\documentclass[journal]{IEEEtran}
\ifCLASSINFOpdf
\else
\fi
\usepackage{amsmath}
\usepackage{dingbat}
\usepackage{blindtext}
\usepackage{array}
\usepackage{graphicx}
\usepackage{balance}
    \usepackage{subfigure}
    \usepackage{subfloat}
    \usepackage{float}
\usepackage{booktabs}% http://ctan.org/pkg/booktabs
\newcommand{\tabitem}{~~\llap{\textbullet}~~}
\usepackage{newunicodechar}
\usepackage{xcolor,colortbl}
\definecolor{Gray}{gray}{0.85}
\definecolor{LightCyan}{rgb}{0.88,1,1}
\usepackage{booktabs}
\newcommand{\xmark}{\ding{55}}%
\usepackage{pifont}
\newcolumntype{a}{>{\columncolor{Gray}}c}
\newcolumntype{b}{>{\columncolor{white}}c}
\usepackage{tikz}
\usepackage{supertabular}
\newcommand{\bluecheck}{}%
\DeclareRobustCommand{\bluecheck}{%
  \tikz\fill[scale=0.5, color=blue]
  (0,.35) -- (.25,0) -- (1,.7) -- (.25,.15) -- cycle;%
}
\usepackage{titlesec}

\begin{document}

	\title{A Survey of Application of  Machine Learning in  Wireless Indoor Positioning Systems }
	
	\author{Amala Sonny,  Abhinav Kumar, \IEEEmembership{Senior Member, IEEE}, and Linga Reddy Cenkeramaddi, \IEEEmembership{Senior Member, IEEE}
 \thanks{ This work was supported in part by the  Department of Science and Technology (DST), Government of India, under Indo-Norway Joint Project. The work of Dr. Abhinav Kumar was also supported in part by TiHAN Faculty Fellowship.}
 
  \thanks{Amala Sonny, and  Abhinav Kumar are with the Department
of Electrical Engineering, Indian Institute of Technology, Hyderabad, Sangareddy, Telengana, 502284, India  (e-mail: ee18resch11009@iith.ac.in; abhinavkumar@ee.iith.ac.in).}
\thanks{Linga Reddy Cenkeramaddi is with the ACPS Group, Department of ICT, University of Agder,
4879 Grimstad, Norway (e-mail: linga.cenkeramaddi@uia.no).}}
% <-this % stops a space

\markboth{Journal of \LaTeX\ Class Files,~Vol.~14, No.~8, August~2021}%
{Shell \MakeLowercase{\textit{et al.}}: A Survey of Application of ML in Wireless Indoor
Positioning Systems}

	\maketitle

	\begin{abstract}

Indoor human positioning  has become increasingly important for applications such as health monitoring, breath monitoring, human identification, safety and rescue operations, and security surveillance. However, achieving robust indoor human positioning remains challenging due to various constraints. Numerous attempts have been made in the literature to develop efficient indoor positioning systems (IPSs), with a growing focus on machine learning (ML) based techniques. This paper aims to compare and analyze current ML-based wireless techniques and approaches for indoor positioning, providing a comprehensive review of enabling technologies for human detection, positioning, and activity recognition. The study explores different input measurement data, including RSSI, TDOA, etc., for various IPSs. Key positioning techniques such as RSSI-based fingerprinting, Angle-based, and Time-based approaches are examined in conjunction with various ML methods. The survey compares the positioning accuracy, scalability, and algorithm complexity, with the goal of determining the suitable technology in various  services.  Finally, the paper compares distinct datasets focused on indoor localization, which have been published using diverse technologies.  Overall, the paper presents a comprehensive comparison of existing techniques and localization models.

	\end{abstract}
	

	\IEEEpeerreviewmaketitle

	\section{Introduction}
	Indoor human detection and positioning have been gaining much attention in recent times. Indoor positioning systems have the objective of determining the real-time location of users or devices within buildings, employing a range of technologies.  The demand for indoor positioning-based military and commercial applications spurred the development of this domain immensely.  It is empowering the Internet of Things (IoT)  by introducing various applications for personal navigation, security and context awareness \cite{intro9}.  IoT integrates rapidly into human life in different ways such as health monitoring for elderly people, activity recognition in smart homes, breath monitoring systems, human identification and detection in safety and rescue operations, security surveillance, and for intrusion detection systems \cite{intro12}. 

 The lack of Line-of-Sight (LoS) between satellites and the receivers renders the Global Positioning System (GPS) and Global Navigation Satellite System (GLONASS) unsuitable for indoor accuracy. Poor indoor connectivity has spurred research for alternative localization methods. Various signals, including Wi-Fi, Bluetooth \cite{bluetooth3}, FM radio \cite{intro8}, and radio-frequency identification (RFID) \cite{RFID4}, are being explored for this purpose. Systems employing Received Signal Strength Indicator (RSSI), OFDM channel state information (CSI), antenna arrays, RFID tags, Wi-Fi fingerprinting, or visible LED lights have achieved sub-meter accuracy, albeit requiring users to carry a wireless device as an access point (AP) \cite{Wi-Fi1}. Fingerprinting techniques are prevalent, involving offline training to generate a radio map and online estimation for location detection using RSS data \cite{wlan1}. However, device heterogeneity may introduce anomalies and  challenging a robust indoor positioning \cite{surv5, Wi-Fi2}.
	
 	Among the existing indoor positioning technologies, vision and sensor network-based approaches face limitations due to privacy concerns, complexity, and the necessity for extensive device coverage. Various systems utilize RADAR  \cite{Rad1}, SONAR \cite{sonar1}, lasers, dedicated sensor networks \cite{wsn1}, and cameras \cite{fmcw3} for indoor positioning and activity recognition. Vision-based systems, reliant on high-resolution cameras and line-of-sight (LoS) conditions, are constrained by illumination requirements and privacy concerns \cite{intro3}. Wearable sensor-based methods necessitate specialized hardware, potentially inconveniencing users, especially the elderly and children, with high installation and maintenance costs  \cite{intro4}. Commercial examples like Xbox Kinect \cite{intro5} utilize depth sensing and vision for motion recognition. However, RADAR and SONAR-based systems primarily detect motion direction via Doppler shift, limiting their applicability in real-life settings due to existing drawbacks in performance and practicality.
\begin{figure*}[!t]
  \centering
		\includegraphics[width=13cm,height=4.5cm]{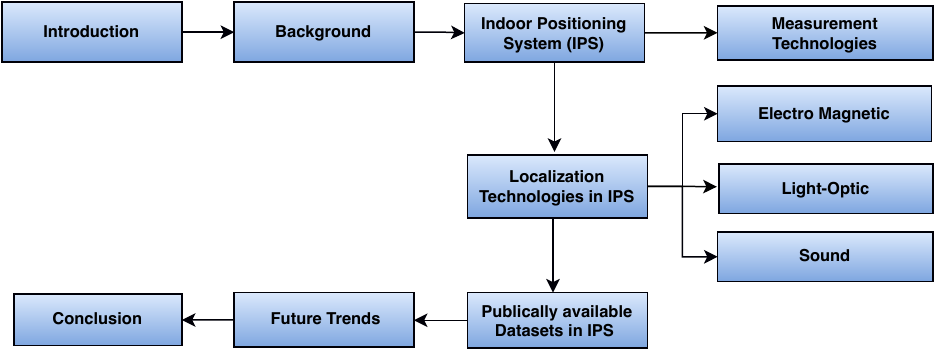}

		\centering \caption{Organization of the Paper. }\label{org}
	\end{figure*}

  	Thus, wireless signals-based positioning, activity sensing, and recognition serve as complementary techniques due to the widespread deployment of wireless devices \cite{intro13}. This passive sensing method, attracting increasing interest, offers device-free detection without privacy concerns, applicable for indoor positioning, gesture recognition, activity recognition, and motion tracking \cite{gesture1, Wi-Fi3, RF1}. Its versatility extends to smart homes, healthcare, security, safety systems, and entertainment applications. Lots of techniques have been proposed based on this using a set of receivers or USRP SDR devices. However, device-free human detection and activity recognition is a very challenging task. Recent studies demonstrate that human presence and activities can significantly affect wireless signals, with multipath signals reflecting, refracting, and scattering. However, environmental factors like multipath fading, non-line-of-sight (NLoS), obstructions, crowd density, and weather conditions complicate the wireless signal in indoor environment. Selecting appropriate mathematical models for signal analysis and target localization is crucial, with triangulation, fingerprinting, and multilateration being common techniques \cite{intro14}. Machine learning (ML) tools, such as multi-layer perceptron (MLP), radial basis function (RBF), k-nearest neighbors algorithm (k-NN), and recurrent neural network (RNN), address non-linearities and enhance data analysis accuracy. Various ML models have been employed in indoor positioning systems, utilizing signals from multiple access points (APs)  \cite{bluetooth3} in wireless sensor networks (WSNs) \cite{intro16,intro17} and neural networks for fingerprint-based approaches, improving positioning performance in WLAN environments.

 Therefore, this survey paper offers a comprehensive examination of current ML-based approaches in wireless indoor positioning. It seeks to provide an in-depth comparison of these technologies, covering various ML methodologies in terms of accuracy, cost-effectiveness, and more. Additionally, the paper explores  the wide array of localization technologies applied in indoor positioning, and publicly available datasets relevant to indoor positioning. Since a combined survey covering all these subdomains is not available, this paper can act as a guide for referencing the state-of-the-art approaches in this emerging area.  The primary motivation behind this work is to shed light on this intricate domain and address research challenges within it. Our goal is to furnish readers with essential references to deepen their understanding and foster further contributions in this area. The paper's organization is outlined in Fig. \ref{org}.

 The structure of the paper is as follows: Sections II and III give insights into the background and Indoor Positioning Systems (IPSs), respectively. Section IV outlines various Machine Learning (ML) techniques employed in diverse IPSs. The comparison and analysis of various positioning technologies are presented in Section V. Section VI showcases publicly accessible datasets in IPSs employing different technologies. Lastly, Section VII outlines forthcoming trends in IPS.
% \vspace{-0.2cm}
 
 \section{Background}

	\begin{table*}
 \renewcommand{\arraystretch}{1}	
 \centering
		\caption{Comparison of existing survey papers in Indoor positioning.}
			\begin{tabular}{ | m{1.9cm} | m{0.5cm}| m{0.6cm} |  m{0.5cm} |  m{0.4cm}| m{0.8cm} | m{0.6cm} |  m{0.8cm} |  m{0.9cm}| m{0.5cm} |  m{0.5cm} |  m{0.6cm}| m{0.6cm} |m{0.5cm}| m{0.5cm}|m{0.6cm}|} 
			\rowcolor{lightgray}

Title &  Year  & ML-based  & VLC  &IR & LiDAR & Sound  & RADAR & Magnetic Sensor & Wi-Fi & BLE & RFID  & UWB & WSN & LTE &  Zigbee \\ \hline

 This Work & 2023 & \bluecheck&\bluecheck&\bluecheck&\bluecheck&\bluecheck&\bluecheck&\bluecheck&\bluecheck&\bluecheck&\bluecheck&\bluecheck&\bluecheck&\bluecheck  &\bluecheck\\ \hline

A. S. Yaro\textit{ et al.} \cite{survnew} & 2023 & \bluecheck &\xmark&\xmark&\xmark&\xmark&\xmark&\xmark&\bluecheck&\xmark&\xmark&\xmark&\xmark&\xmark &\xmark \\ \hline

A. Shastri \textit{ et al.} \cite{fmcw3} &2022 &\xmark& \xmark &\xmark&\xmark&\bluecheck&\xmark&\xmark&\xmark&\xmark&\xmark&\xmark&\xmark&\xmark&  \xmark  \\ \hline

P. S. Farahsari\textit{ et al.} \cite{surv10}  & 2022 & \bluecheck &\bluecheck&\xmark&\xmark&\bluecheck&\xmark&\xmark&\bluecheck&\bluecheck&\bluecheck&\bluecheck&\xmark&\bluecheck&\bluecheck\\ \hline

 M. Elsanhoury\textit{ et al.} \cite{surv14}&2022& \bluecheck &\xmark&\xmark&\xmark&\xmark&\xmark&\xmark&\xmark&\xmark&\xmark&\bluecheck&\xmark & \xmark &\xmark\\ \hline

P. Roy $\&$ C. Chowdhury \cite{survnew2}& 2022&\bluecheck &\xmark&\xmark&\xmark&\xmark&\xmark&\xmark&\bluecheck&\xmark&\xmark&\xmark&\xmark &\xmark &\xmark \\ \hline

  Yang\textit{ et al.} \cite{surv9} &2021 &\xmark &\xmark&\xmark&\xmark&\bluecheck&\xmark&\xmark&\bluecheck&\bluecheck&\bluecheck&\bluecheck&\xmark&\xmark &\bluecheck\\ \hline

 F. Alam\textit{ et al.}\cite{survnew8} &2021 & \bluecheck &\bluecheck&\bluecheck&\xmark&\xmark&\xmark&\xmark&\xmark&\xmark&\xmark&\xmark&\xmark&\xmark & \xmark \\  \hline

 P. Roy $\&$ C. Chowdhury\cite{survnew9} &2021 & \bluecheck &\xmark&\xmark&\xmark&\bluecheck&\bluecheck&\bluecheck&\bluecheck&\bluecheck&\bluecheck&\xmark&\xmark&\xmark&\xmark\\  \hline

N. Singh\textit{ et al.}\cite{survnew4} & 2021 & \bluecheck&\xmark&\xmark&\xmark&\xmark&\xmark&\xmark&\bluecheck&\xmark&\xmark&\xmark&\xmark & \xmark &\xmark\\  \hline

A. Nessa\textit{ et al.} \cite{survnew3}& 2020 & \bluecheck&\xmark&\xmark&\xmark&\xmark&\xmark&\xmark&\bluecheck&\bluecheck&\bluecheck&\bluecheck&\xmark&\xmark &\bluecheck\\ \hline

 F. Zafari\textit{ et al.} \cite{surv6}  &2019 & \xmark&\bluecheck&\xmark&\xmark&\bluecheck&\xmark&\xmark&\bluecheck&\bluecheck&\bluecheck&\bluecheck&\bluecheck& \xmark&\bluecheck\\ \hline
   
 N. Saeed\textit{ et al.} \cite{surv11} & 2019 & \xmark &\xmark&\xmark&\xmark&\xmark&\xmark&\xmark&\xmark&\xmark&\xmark&\xmark&\bluecheck & \xmark  &\xmark\\ \hline

F. Khelifi\textit{ et al.} \cite{survnew5} & 2019 & \xmark&\xmark&\xmark&\xmark&\xmark&\xmark&\xmark&\xmark&\xmark&\xmark&\xmark&\bluecheck & \xmark &\xmark \\ \hline

  W. Liu \textit{ et al.} \cite{surv13} &2019 & \bluecheck &\xmark&\xmark&\xmark&\xmark&\xmark&\xmark&\bluecheck&\xmark&\xmark&\xmark&\xmark & \xmark &\xmark\\ \hline

R. F. Brena \textit{ et al.} \cite{survnew6} & 2017 & \xmark &\bluecheck&\bluecheck&\xmark&\bluecheck&\xmark&\bluecheck&\bluecheck&\xmark&\bluecheck&\xmark&\bluecheck&\xmark &\bluecheck\\ \hline 

S. Xia\textit{ et al.}\cite{survnew10}& 2017 & \bluecheck &\xmark&\xmark&\xmark&\xmark&\xmark&\xmark&\bluecheck&\xmark&\xmark&\xmark&\xmark&\xmark& \xmark\\ \hline

 A.Khalajmehrabadi \textit{ et al.} \cite{intro13} &2017 &\xmark &\xmark &\xmark &\xmark &\xmark &\xmark &\xmark &\bluecheck&\xmark &\xmark &\xmark &\xmark &\xmark  &\xmark \\ \hline

	A. K. Paul $\&$ T. Sato\cite{localization2}&2017 & \xmark&\xmark&\xmark&\xmark&\xmark&\xmark&\xmark&\xmark&\xmark&\xmark&\xmark&\bluecheck & \xmark  &\xmark\\ \hline

  J.Xiao\textit{ et al.} \cite{surv7} &2016&\bluecheck& \xmark& \xmark& \xmark& \bluecheck&\xmark&\bluecheck&\bluecheck& \bluecheck & \bluecheck & \bluecheck & \xmark & \bluecheck & \bluecheck\\ \hline

A. M. Ossain $ \&$ W.-S. Soh  \cite{surv5}  &2015 & \xmark&\xmark&\bluecheck&\xmark&\xmark&\bluecheck&\xmark&\bluecheck&\bluecheck&\bluecheck&\bluecheck&\xmark&\xmark&  \xmark\\ \hline

S.He $\&$ S.-H. G. Chan \cite{Wi-Fi2} &2015 & \xmark&\xmark&\xmark&\xmark&\xmark&\xmark&\xmark&\bluecheck&\xmark&\xmark&\xmark&\xmark&\xmark &\xmark\\ \hline

M.A. Al-Ammar \textit{ et al.} \cite{surv1} &2014 & \xmark&\xmark&\xmark&\xmark&\xmark&\xmark&\xmark&\bluecheck&\xmark&\xmark&\xmark&\xmark & \xmark&\xmark \\ \hline

F. Ijaz\textit{ et al.} \cite{ultrasonic} &2013&\xmark&\xmark&\xmark&\xmark&\bluecheck&\xmark&\xmark&\xmark&\xmark&\xmark&\xmark&\xmark & \xmark &\xmark\\ \hline

 K. Al Nuaimi $\&$ H. Kamel \cite{surv4} &2011 & \xmark&\bluecheck&\bluecheck&\xmark&\bluecheck&\bluecheck&\xmark&\xmark&\xmark&\xmark&\bluecheck&\xmark&\xmark &\xmark\\ \hline
 
   % R. Mautz $\&$  S. Tilch \cite{optic1} & 2011&\xmark&&&&&&&&&&&&& \\ \hline

   T. Teixeira\textit{ et al.}  \cite{survnew7} & 2010 & \xmark &\xmark&\bluecheck&\bluecheck&\bluecheck&\bluecheck&\xmark&\bluecheck&\xmark&\bluecheck&\bluecheck& \xmark&\xmark& \xmark\\ \hline
   
Y. Gu \textit{ et al.} \cite{IPS9} &2009 &\xmark&\xmark&\bluecheck&\xmark&\bluecheck&\bluecheck&\bluecheck&\bluecheck&\bluecheck&\bluecheck&\bluecheck&\xmark& \xmark &\xmark\\ \hline

I. Guvenc $\&$ C.-C.Chong \cite{toa0} &2009& \xmark &\xmark&\xmark&\xmark&\xmark&\xmark&\xmark&\xmark&\xmark&\xmark&\bluecheck&\bluecheck&\bluecheck&\xmark\\ \hline

  H. Liu \textit{ et al.} \cite{tdoa1} & 2007 & \xmark&\xmark&\xmark&\xmark&\xmark&\xmark&\xmark&\bluecheck&\bluecheck&\bluecheck&\bluecheck&\xmark&\bluecheck &\xmark\\ \hline

			\end{tabular}
   \label{survcomp}
   \end{table*}

	\begin{figure*}[h]
    \includegraphics[width=17cm,height=7cm]{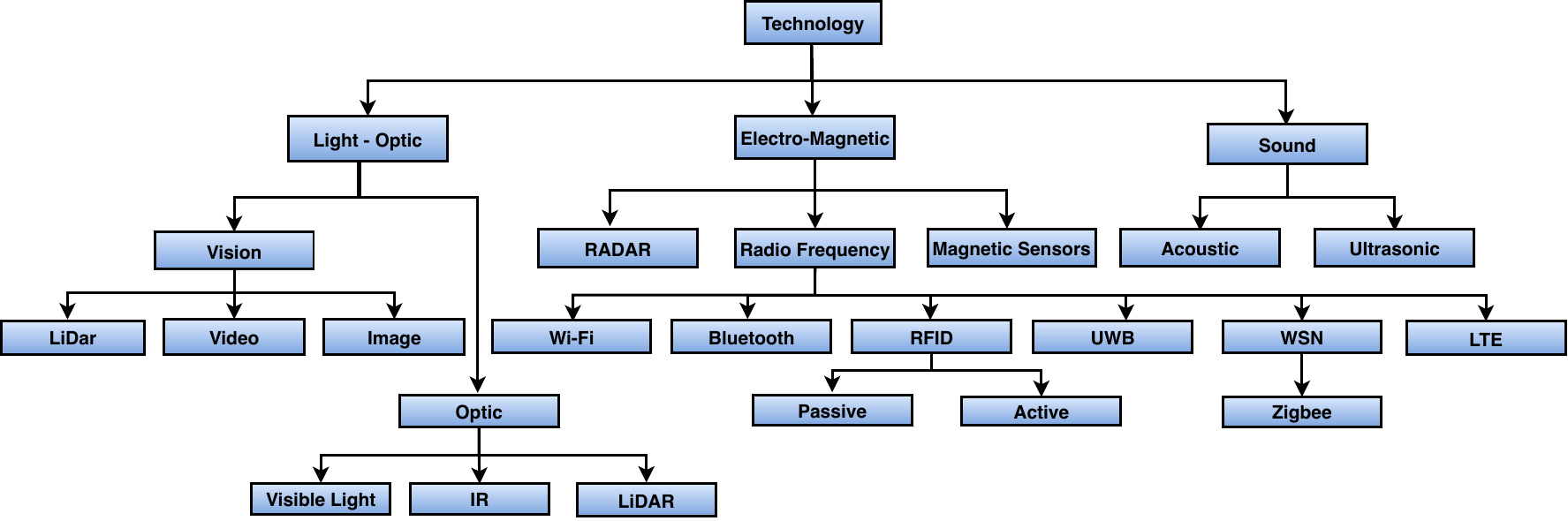}
		\centering \caption{ Classification of IPS-based on positioning technology. Vision-based techniques are not covered in this article. }\label{img2}
	\end{figure*}

Wireless signals encounter multipath effects from reflection, refraction, and scattering in both indoor and outdoor environments, posing challenges for accurate location tracking. Indoor Positioning Systems (IPS) address indoor navigation difficulties using technologies like Wi-Fi, Bluetooth, and UWB, enhancing location-based services, asset tracking, and security in places such as malls, airports, and offices \cite{RF2}. IPS-based systems are crucial for modern indoor navigation, meeting the demand for precise location data and facilitating mapping, optimization, and improvement of indoor spaces. Unlike outdoor environments dominated by GPS, indoor settings face signal attenuation and multipath complexities. IPSs systematically gather data to evaluate scenes and positioning methods, improving accuracy and effectiveness in locations like malls and airports. 

Here, we are comparing some of the most referred IPS-based surveys in the literature in terms of tasks involved, technology adoption, infrastructure deployment, data collection approach, signal analysis, and algorithm testing. Survey outcomes inform IPS deployment for asset tracking, security enhancement, and personalized services \cite{RF3}. Bridging the gap between technology and practical solutions, IPS surveys are pivotal in the evolving indoor landscape.

The papers \cite{Wi-Fi2, intro13, survnew2, survnew, survnew4, surv13, surv1} and \cite{survnew10} are  IPS-based surveys, with an exclusive focus on Wi-Fi-based methods. The papers \cite{Wi-Fi2, intro13, survnew10} discuss the popularity of indoor location-based services (ILBS) and the use of Wi-Fi (802.11) as a promising signal for indoor positioning, particularly Wi-Fi fingerprinting. Meanwhile,  \cite{survnew2,surv1} explores indoor positioning technologies, techniques, and algorithms, challenges, proposed solutions, categorization, and potential enhancements for accurate location tracking.  

The surveys in \cite{survnew, survnew4} identify the factors involving offline (RSS map creation) and online (user positioning) phases,  factors affecting performance, discusses their impact, mitigation strategies, and future trends in 2D RSS fingerprinting-based I-WLS. Additionally, \cite{surv13} provides a comprehensive review of indoor positioning methods based on Wi-Fi's CSI, comparing geometry-based and fingerprint-based positioning systems and discussing multi-source fusion technologies and the transition from traditional ML to deep learning methods.

The papers \cite{surv11, survnew5, localization2} delve into Wireless Sensor Networks (WSNs)-based positioning, covering both range-based and range-free methods. They explore measurement techniques, applications, and challenges like accuracy, cost, complexity, and scalability. Survey \cite{surv14} focuses on Ultra-WideBand (UWB) positioning systems, highlighting advantages for indoor positioning and smart logistics.  The study in \cite{ultrasonic} compares real-time indoor location systems, emphasizing RF and ultrasonic technologies, and highlighting ultrasonic systems' fine-grained location capabilities at low costs. Similarly, \cite{fmcw3} reviews mmWave communication and radar devices for indoor positioning and sensing, emphasizing their potential in 5G and 6G networks, stressing the need for improved hardware and integration. Passive RF or wireless sensing approaches in IPSs are unobtrusive and well-researched. The survey in \cite{survnew8} reviews non-RF techniques like visible light, infrared, and electric field sensing, discussing limitations and future directions.  And \cite{toa0} offers a unified survey of Time-of-Arrival (TOA)-based positioning and NLoS mitigation algorithms, covering bounds, estimators, NLoS techniques, challenges, and future research. This comprehensive review explores wireless device localization through TOA from base stations, encompassing various algorithms with varying accuracies, complexities, and robustness against NLoS effects.

The surveys conducted by \cite{surv9, surv6} provide an extensive overview of indoor positioning techniques encompassing Angle of Arrival (AoA), Time of Arrival (ToF), Round trip Time of Flight (RToF), and Received Signal Strength (RSS), utilizing technologies like Wi-Fi, RFID, UWB, and Bluetooth. These surveys emphasize human users and devices, spotlighting system strengths, and assessing aspects such as energy efficiency, cost, and accuracy. In particular, \cite{surv9} delves into supervised (SVM, KNN, NN) and unsupervised (isolation forest, k-means, EM) approaches, as well as Bayesian filtering methods.  In a similar vein, \cite{ surv5, surv12, surv4, survnew6, survnew7} explore emerging solutions and introduce additional evaluation criteria, assessing various proposed positioning systems through both traditional and novel criteria. Notably, \cite{survnew7} compares active vs. passive sensors, single-modality vs. sensor fusion approaches, and instrumented vs. non-instrumented settings.
Wireless IPSs have gained traction in applications like asset tracking and inventory management. In this context, \cite{IPS9, tdoa1} furnish an overview of existing wireless technology-based solutions, categorizing techniques and systems. The focus lies on analyzing triangulation, scene analysis, and proximity, with a special emphasis on location fingerprinting. Furthermore, the study presented in \cite{IPS9} delves into personal networks (PNs) that interconnect users' devices via various communication technologies to offer location-aware services.

The articles  \cite{surv10, survnew9, survnew3,  surv7} provide an extensive overview of diverse indoor positioning methods utilizing technologies such as Visible Light Communication (VLC), Infrared (IR), Sound, Wi-Fi, and Bluetooth Low Energy (BLE) etc., The focus is on machine learning-based techniques employed by these approaches to create IPSs. The referenced paper offers a detailed exploration of machine learning methods for indoor positioning using technologies like Wi-Fi and Bluetooth. It examines traditional algorithms, feature selection, and emerging Deep Learning techniques, addressing challenges and advancements in the field.

The paper covers the applicability of machine learning, basic principles, challenges, and recent advancements, offering insights into the future of indoor positioning research. 
The survey in \cite{surv10, surv7} presents a comprehensive survey of wireless indoor localization from a device perspective, reviewing recent advances in both modes, emphasizing smartphone integration and human-centric device-free approaches. It compares schemes based on accuracy, cost, scalability, and energy efficiency, and addresses intrinsic challenges and open research issues in the field. The discussion in \cite{surv10} extends to location-based services (LBSs), real IoT applications, and notable vendors, offering valuable insights for future IoT-driven research in positioning services.
\par A comprehensive technological comparison of the surveys discussed is presented in Table. \ref{survcomp}. The analysis indicates that this study excels in terms of encompassing a wide range of technologies associated with IPSs.

 \section{Indoor Positioning System (IPS)}
Indoor positioning systems (IPS) aim to determine the location of users or devices within buildings using diverse technologies in real-time. As defined in \cite{IPS2}, an IPS must furnish the real-time location of entities within physical environments like gyms, hospitals, and shopping malls. Ideally, such systems should swiftly update target locations while covering vast areas. The data offered by IPSs varies based on user requirements; some provide absolute target locations by referencing predefined areas, crucial for precise tracking and navigation. Others offer relative position information, detecting the status of different target components. Proximity location, estimating approximate target locations, is another data type obtainable from IPSs. For instance, IPSs may indicate the room where a patient resides in a hospital. Each type serves distinct purposes, catering to diverse applications and user needs.

IPSs rely on a diverse set of components to accurately determine location within indoor environments. Sensing technology, including Wi-Fi, Bluetooth, and RFID sensors, forms the foundational elements for gathering device positions and signals \cite{IPS3, IPS4}. Various IPSs employ these technologies independently or in combination to sense the environment and attribute data accordingly. The classification of IPSs based on positioning technology is outlined in Fig. \ref{img2}. In this survey, we are covering all the technologies mentioned in Fig. \ref{img2} except the vision-based approaches considering the length of the study. Each technology offers distinct advantages and limitations, with selection based on the service requirements of the IPS.

\begin{figure}[!tp]
		\includegraphics[width=9cm,height=3cm]{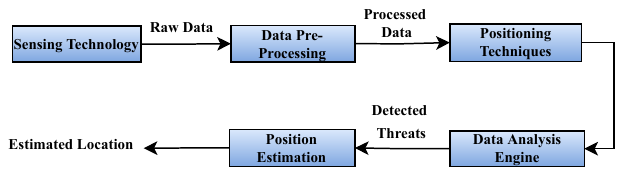}
		\centering \caption{Components of an Indoor Positioning System.}\label{IPS}
  \label{IPSblocks}
	\end{figure}

Data preprocessing plays a crucial role in refining raw sensor data for accurate positioning calculations. Techniques such as noise reduction, data fusion, signal strength calibration, and map matching are essential for enhancing accuracy and reliability \cite{IPS5, prepro1, prepro2}. Moreover, outlier detection methods and data privacy measures are pivotal for robust positioning while ensuring user privacy \cite{prepro4, prepro5}.  However, preprocessing aspects related to IPSs are not within the scope of this survey. Ultimately, a robust data analysis engine, powered by algorithms and machine learning, refines accuracy and updates real-time positions, forming the framework for IPSs. These interconnected components enable a variety of applications, including navigation services and asset tracking. A visual representation of the typical IPS building blocks is provided in Fig. \ref{IPSblocks}.

% \vspace{-0.2cm}

 \subsection{Positioning Techniques}
 In this section, we explore various indoor positioning methods used to determine the location of objects within indoor environments. We will discuss each of the main methods in detail.

	\subsubsection{Triangulation}
 
 RSSI, Angle of Arrival (AoA) and Time of Arrival (TOA) are  methods for  target position calculation by using the geometric properties of triangles \cite{triang1, toa1}. Among these, TOA is the most accurate and complex and is  unaffected by  Multipath effects  in indoor scenarios. RSS and TOA require at least 3 reference nodes (RNs) for localization. The basic principle of localizing a target in 2D or 3D position using TOA or RSS is given in Fig. \ref{triang}.  Localization in 2D or 3D space via TOA or RSS depends on knowing the positions of 2 or 3 RNs respectively. The method estimates the transmitter's location based on angles and distances between the RNs, improving accuracy with more RNs particularly effective with Line of Sight (LoS) conditions. Bluetooth 5.1 uses a similar approach for its positioning system. The three RNs - RN1, RN2, RN3 have geographical coordinates at (x1,y1), (x2,y2), (x3,y3) respectively. The exact location of the target at $(x,y)$ can be estimated using the angles $(\alpha, \beta, \gamma ) $ and
the distance between RNs in Fig. \ref{triang}. AOA is less complex and it requires only two reference points. However, it may have less accuracy if the target is far away.

\begin{figure*}[!ht]
\centering
\subfigure{\includegraphics[width=5.8cm,height=3.8cm]{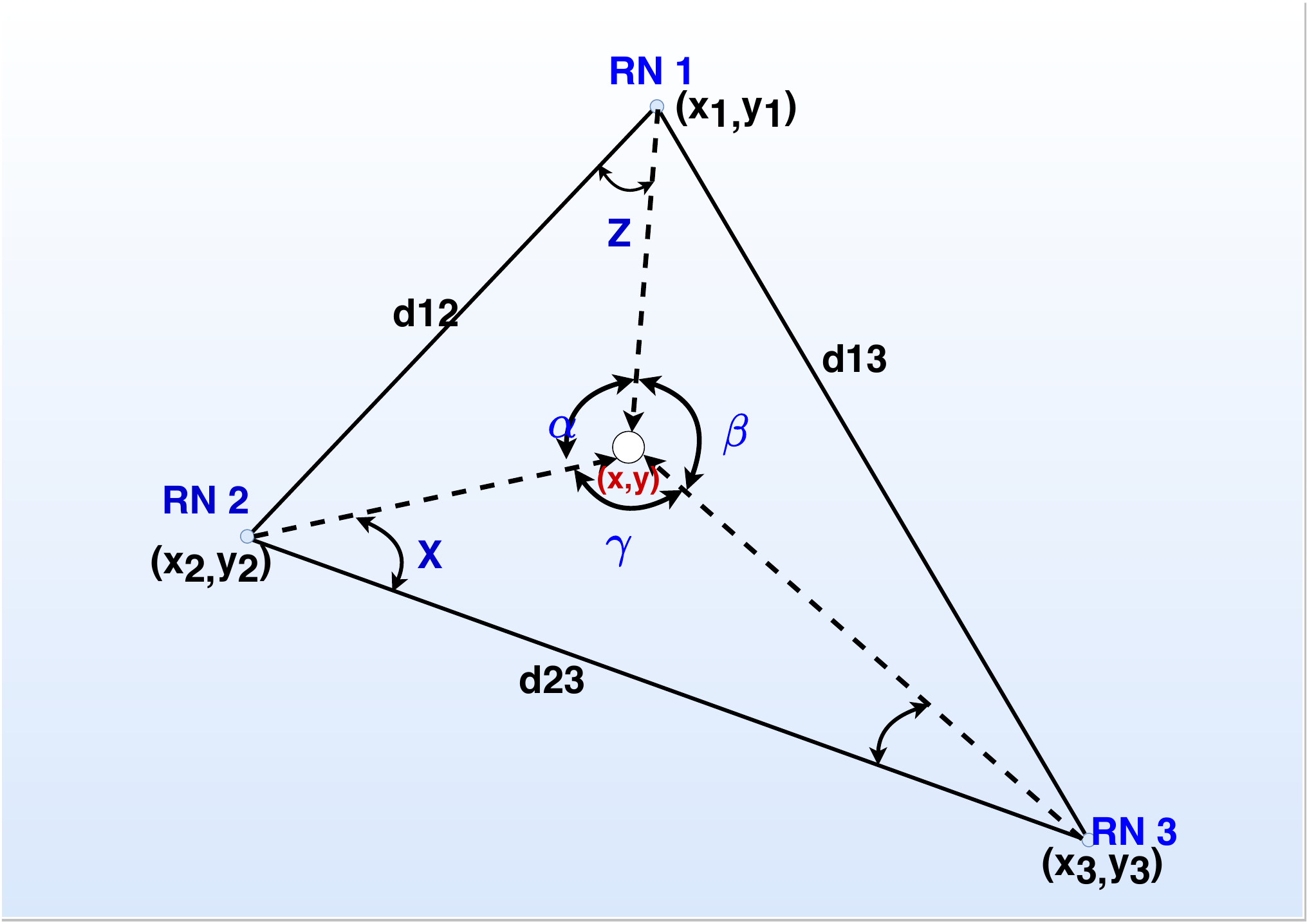}\label{triang}}
\subfigure{\includegraphics[width=5.8cm,height=3.8cm]{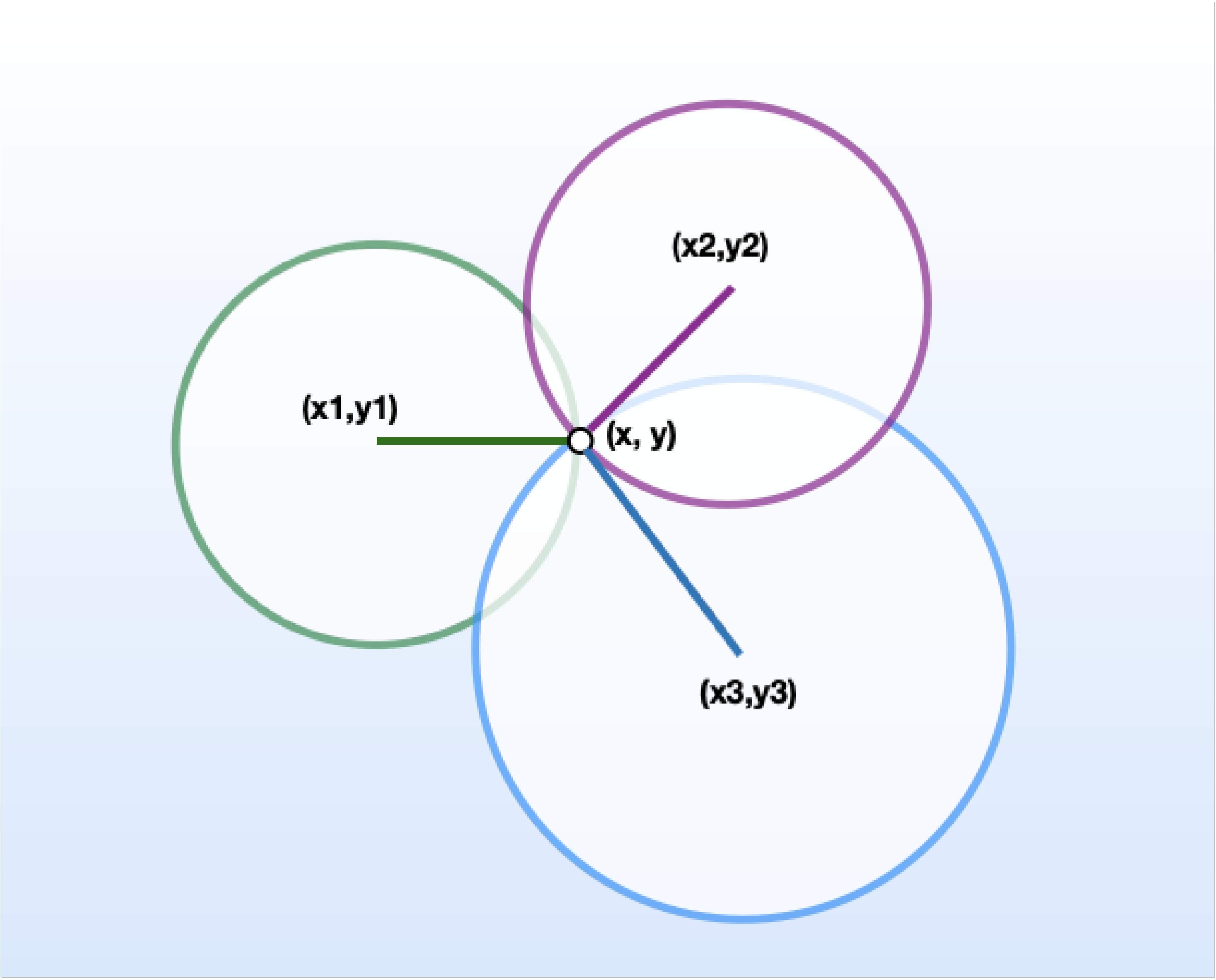}\label{trilat}}
\subfigure{\includegraphics[width=5.8cm,height=3.8cm]{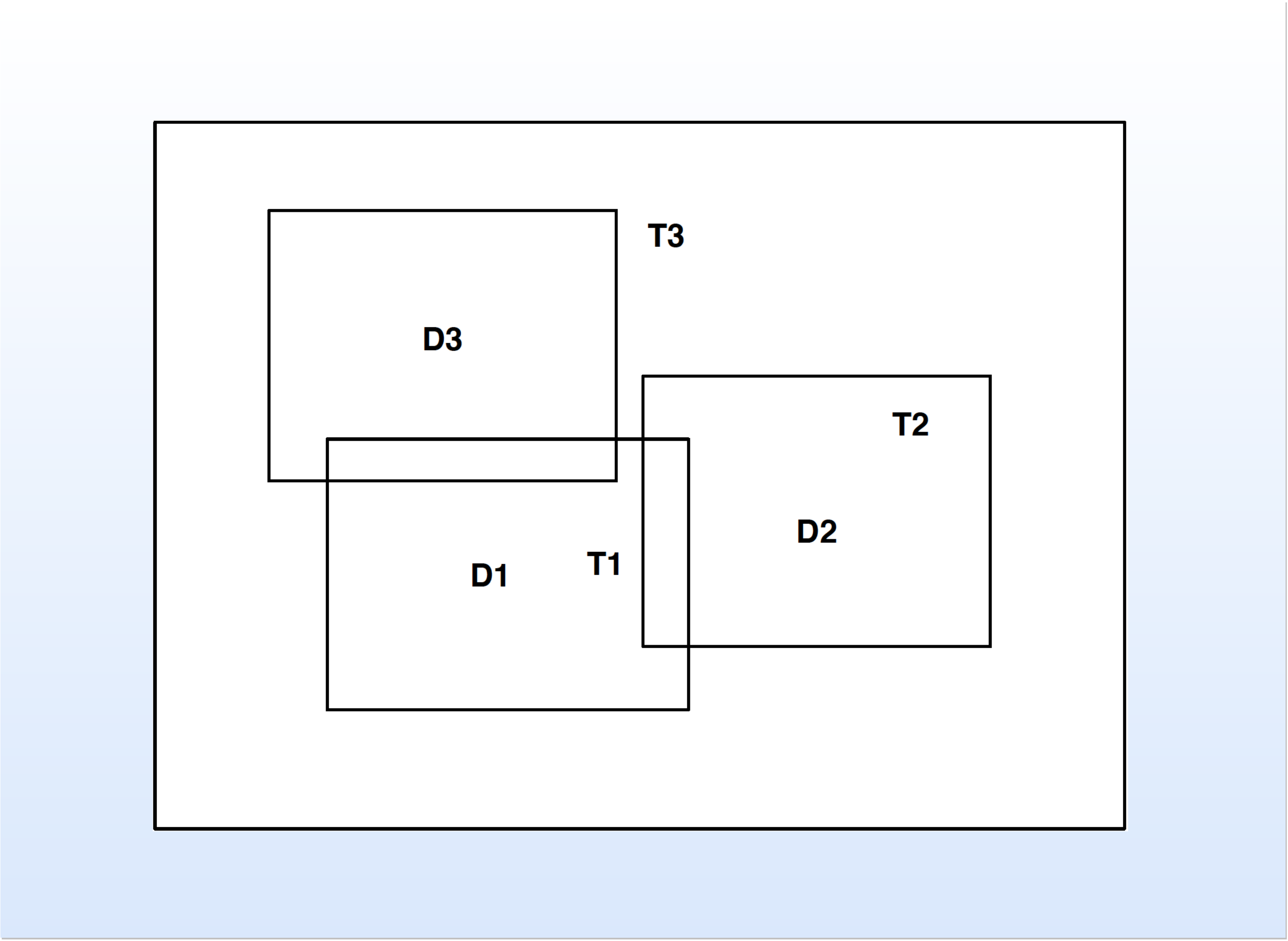}\label{prox}}\\
a) Tringulation  \hspace{3.4cm} b) Trilateration  \hspace{3.5cm} c) Proximity\\
\caption{ Positioning Techniques}
% \label{buildingcla}
% \vspace{-0.5cm}
\end{figure*}

\subsubsection{Trilateration}

rilateration, a technique for determining the position of objects in 2D or 3D space using distances from known reference nodes (RNs), calculates intersections of circles (2D) or spheres (3D) centered at these points. It is widely used in GPS, indoor positioning, and robotics, along with challenges from measurement errors and signal interference. Increasing the number of reference nodes improves system reliability and accuracy. Fig. \ref{trilat} demonstrates 2-D localization, where the object's position intersects at least three circles/spheres from anchor nodes. System reliability improves with more anchor nodes, aiding accuracy. $(x, y)$ denotes the unknown object's coordinates, while $(x1, y1), (x2, y2), (x3,y3)$ represent known RNs \cite{toa2,toa3}.

	\subsubsection{Fingerprinting}

 Fingerprinting positioning enhances IPS accuracy by incorporating prior location data. It comprises offline mapping using RSSI signals stored in a database and online comparison of current RSSI values to determine the nearest point, thus locating the target. Matching methods encompass deterministic and probabilistic algorithms, with the latter offering superior performance albeit with increased computational demands. Techniques like KNN and SVM provide satisfactory results, while DNNs with SDAE yield precise fingerprinting outcomes. Despite its effectiveness, fingerprinting is laborious, time-consuming, and reliant on predefined maps, necessitating remapping for environmental changes. Various strategies are employed to mitigate these challenges \cite{ ours1, rss13, rss14}.

	\subsubsection{Proximity}
 
In certain situations, pinpointing precise object locations might be unnecessary, with an approximate zone around the subject proving sufficient. Beacons detect objects approaching Points of Interest (PoIs), with the object's detection zone determined by adjusting Tx-power. Proximity location sensing compares the target's position with a known location in the same environment. If a target is detected by any detector within the scenario, its location is deemed within that detector's proximity area \cite{IPS9}. In Fig. \ref{prox}, each square denotes the proximity area of detectors $D1, D2$, and $D3$, tracking targets $T1$ and $T2$ by detectors D1 and D2 respectively. Target T3 remains untracked as it's outside any detector's proximity. While unable to provide absolute target location like other methods, it suffices for estimating target presence in a room. Proximity positioning solely offers proximity data, prompting IPSs to integrate multiple techniques for enhanced accuracy in diverse applications like hospitals and malls.
 	
	\subsubsection{Trajectory}
 Trajectory-based Indoor Positioning Systems (IPS) utilize object or user movement patterns to determine positions in indoor settings. Unlike methods relying solely on instant measurements, trajectory-based IPSs leverage continuous motion data from mobile devices or sensors. These data, including accelerations, rotations, or step counts, undergo analysis using algorithms to estimate object trajectories. Comparing these trajectories with reference data or employing map-matching techniques allows the system to determine the current object position \cite{trajec1}.
 
	\subsection{Sensing Techniques}

This section discusses sensing techniques derived directly from receive signals of the  corresponding networks. Signals propagate through either LoS or NLoS paths indoors. LoS paths allow accurate distance estimation but pose challenges indoors due to obstructions. NLoS paths introduce additional attenuation from obstacles, causing distance estimation errors. Hence, various techniques are necessary for precise localization, as discussed in this section \cite{NLOSnew1}.

\subsubsection{Signal Based}
This method is the most straightforward and frequently employed in indoor localization settings. It relies on measuring the power and strength of the received signals.
\paragraph{ Received Signal Strength (RSS)}
	
	RSS-based localization, integral to Distance-Based Localization (DBL), relies on trilateration. RSSI, indicating the power received by a receiver from the radio signal, is a prevalent data type derived from networks like GSM, WLAN, GPS, and Bluetooth \cite{rss2,rss3,rss4}. It involves using RSS measurements to estimate the distance between the user device and at least three reference points, enabling geometric calculations for spatial coordinates determination, as illustrated in Fig. \ref{rssifig}. Measurement-Based Localization (MBL) similarly utilizes RSS measurements from reference points to ascertain the user device's position, often requiring a central controller or ad-hoc communication network for data processing. Both model-based \cite{rss7,rss8} and fingerprint-based \cite{rss9} localization approaches leverage this data. The model-based method employs a log-normal path loss model to estimate distances to User Equipment (UE) \cite{rss12}. Despite its ubiquity, RSS suffers in indoor scenarios due to fluctuations. While the RSS-driven approach is cost-effective, its accuracy is compromised, especially where line-of-sight is obstructed, leading to decreased precision due to signal attenuation and multipath fading \cite{rss13,rss14}. Although filtering methods can mitigate some issues, achieving high localization accuracy without sophisticated algorithms remains 

	\paragraph{ Channel State Information (CSI)}
	
	CSI is information that represents the estimates of the channel of a communication link. It is a fine-grained measurement of the physical layer properties including the amplitude and the phase of each orthogonal subcarrier in the channel. It describes how a signal is propagating from the transmitter to the receiver and the extent of scattering, fading, and power decay throughout the propagation. Recent studies showed the capability of CSI measurements for localization because of the multipath effects \cite{localization13,localization16}.

\subsubsection{Time Based}
This technique relies on measuring the propagation time of signals. The distance between RNs and the mobile node (MN) is calculated by multiplying the time of propagation with the signal speed.
\paragraph{  Time of Arrival (TOA)}

TOA, denoting the absolute time for a signal to travel from an RN to a UE \cite{toa0}, offers energy-efficient real-time localization. However, precise clock synchronization from at least 3 RNs is necessary for accuracy \cite{toa1}. Assuming the signal's transmission at time 0 from the UE and reception by the i$^{th}$ RN at time $t_i$, $t_i$ serves as TOA. The UE's location is determined by intersecting circles centered at each RN with $d_i$ as radius \cite{toa2,toa3,toa4}. Cheung $ et ~ al.$ proposed an approach for localizing UE in both LoS and NLoS scenarios \cite{toa9}.
 Consequently, the separation between entities $RN_i$ and $UE$  can be determined employing equation
 \begin{equation*}
     d_i=(t_2-t_1) \times c
 \end{equation*}

\begin{figure*}[!ht]
\centering
\subfigure{\includegraphics[width=5.8cm,height=3.8cm]{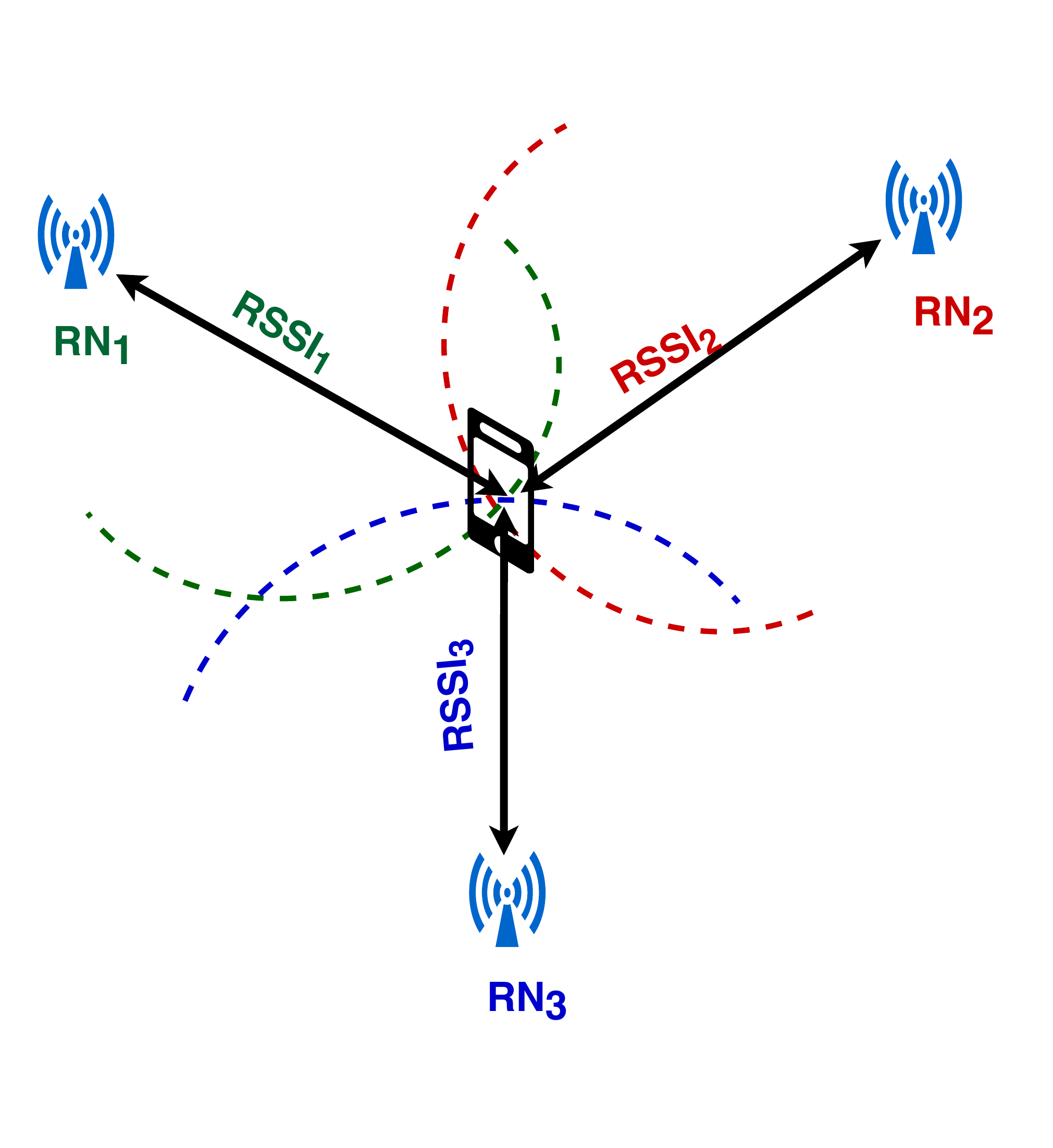}\label{rssifig}}
\subfigure{\includegraphics[width=6cm,height=4.55cm]{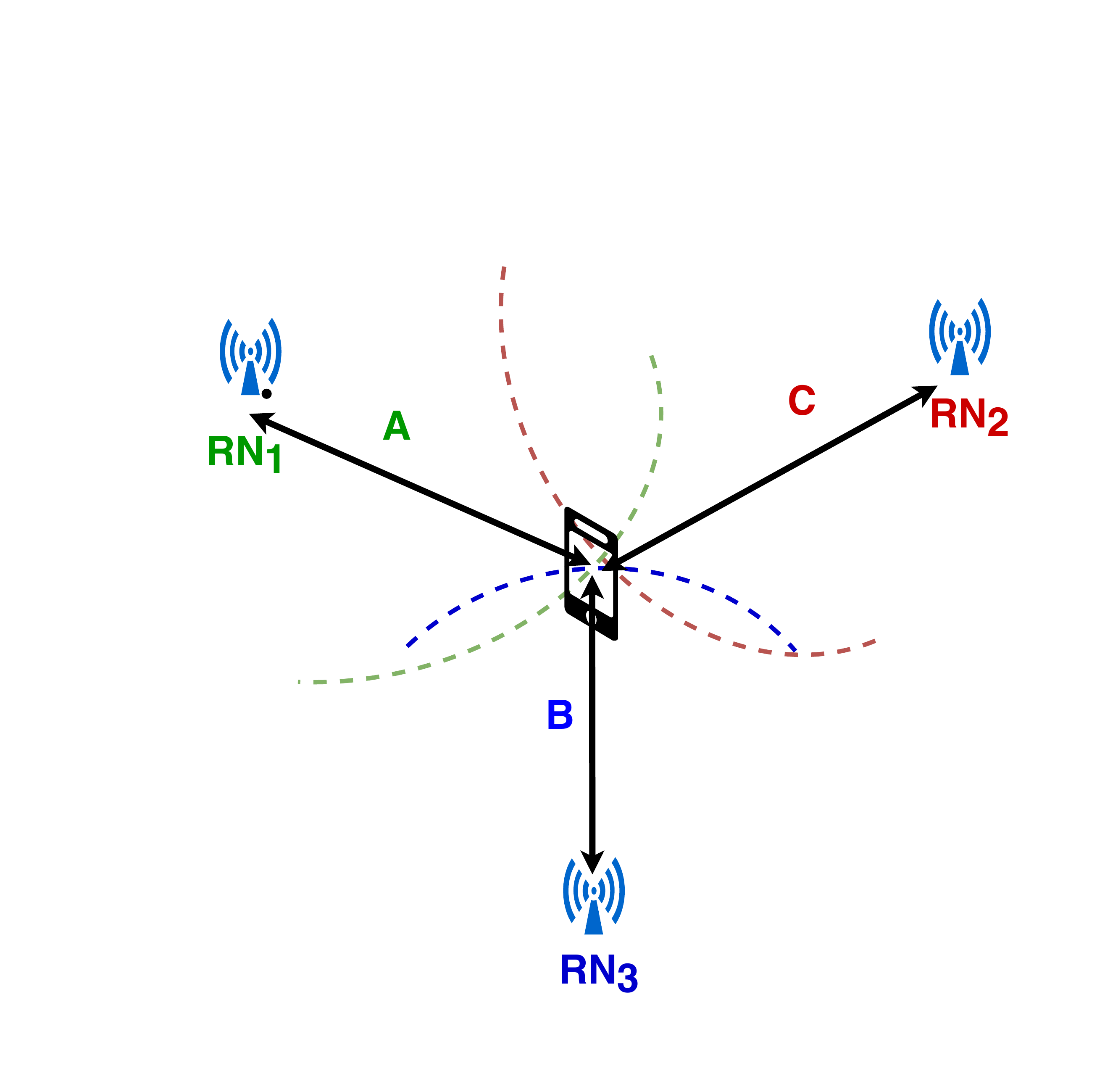}\label{toafig}}
\subfigure{\includegraphics[width=5.8cm,height=4cm]{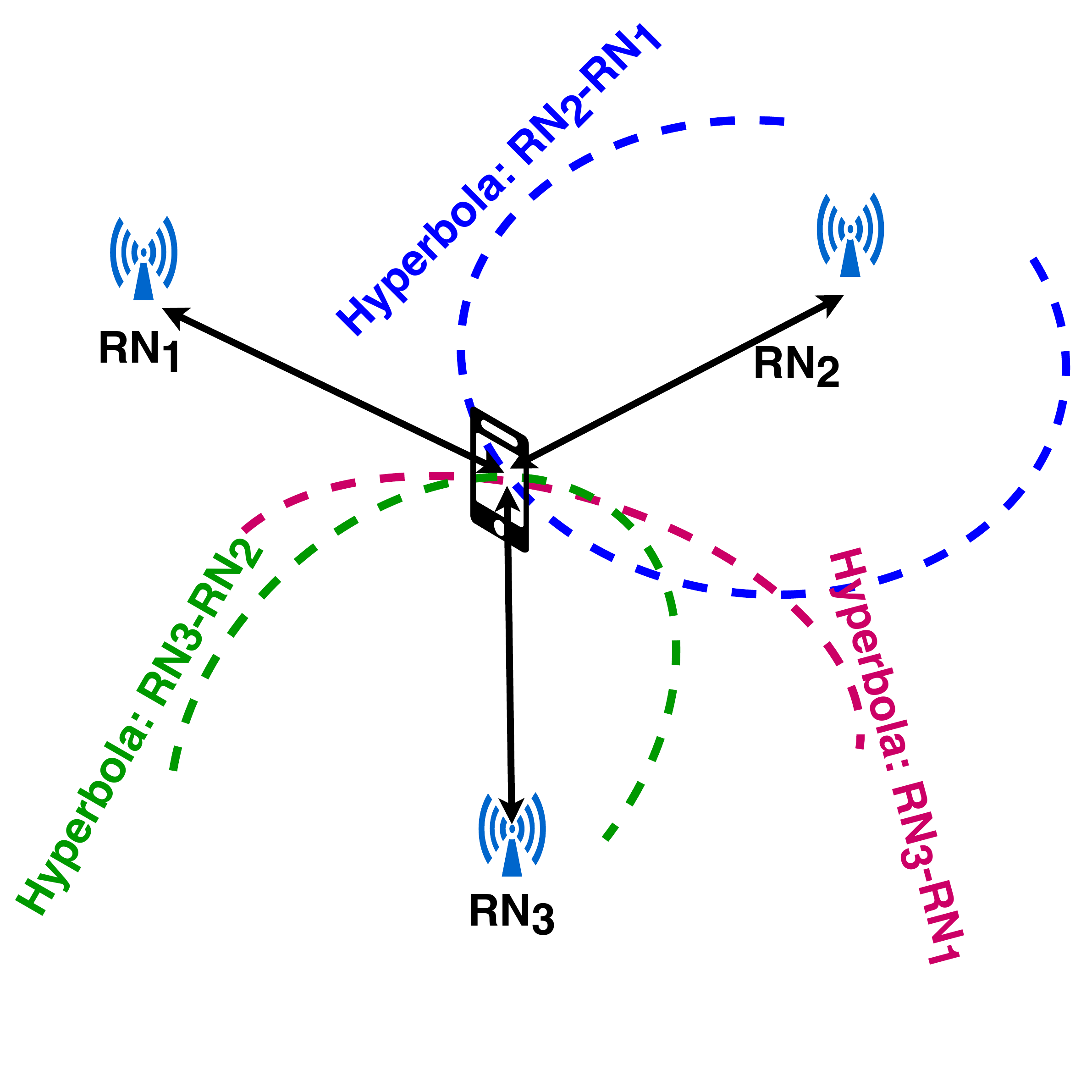}\label{tdoafig}}\\
      a) RSSI  \hspace{4cm} b) TOA  \hspace{4cm} c) TDOA\\
\caption{ Positioning Techniques}
% \label{buildingcla}
\vspace{-0.5cm}
\end{figure*}

	\paragraph{   Time Difference Of Arrival (TDOA)}

The location of the UE is identified by using three or more remote receivers (RNs) capable of detecting signals of interest. Each RN is time-synchronized to capture the reference signal, enabling the calculation of the distance between the UE and RNs based on differences in arrival times. Multiple receivers provide hyperbolic curves as solutions, with the UE located at their intersection \cite{tdoa2,tdoa3,localization15}. The system of hyperbola equations can be resolved through linear regression \cite{tdoa1} or Taylor-series expansion approximation. Fig. \ref{tdoafig} demonstrates the utilization of three distinct Reference Nodes (RNs) to determine the two-dimensional location of any target, illustrating the hyperbolas formed from RN measurements for location determination. Similar to ToA techniques, TDoA precision depends on factors like signal bandwidth, receiver sampling rate, and unobstructed line of sight. Stringent synchronization is crucial, although TDoA requires synchronization only among transmitters, unlike ToA techniques that necessitate synchronization between the transmitter (Tx) and receiver (Rx).
	
	\paragraph{Round trip Time of Flight (RToF) }	
 Round-trip time of Flight (RToF) measures the spatial gap between MN and RNs by timing a signal's journey from MN to RN and back. Similar to time-of-flight (ToF), RToF calculates the full round-trip ToF. RToF requires moderately synchronized clocks between MN and RNs, unlike ToF. However, its precision is influenced by factors like sampling rate and signal bandwidth, more so due to double signal transmission. Latency in receiver response is a concern, especially in short-range systems like indoor localization \cite{rtt,rtt2,rtt3}.

\subsubsection{Angle-based}
Angle-based methods utilize signal angles, such as radio waves or light, to estimate user or device positions by measuring arrival angles from multiple RNs, enabling precise localization.

 \begin{figure}[!t]
\centering
\subfigure{\includegraphics[width=4.35cm,height=3.7cm]{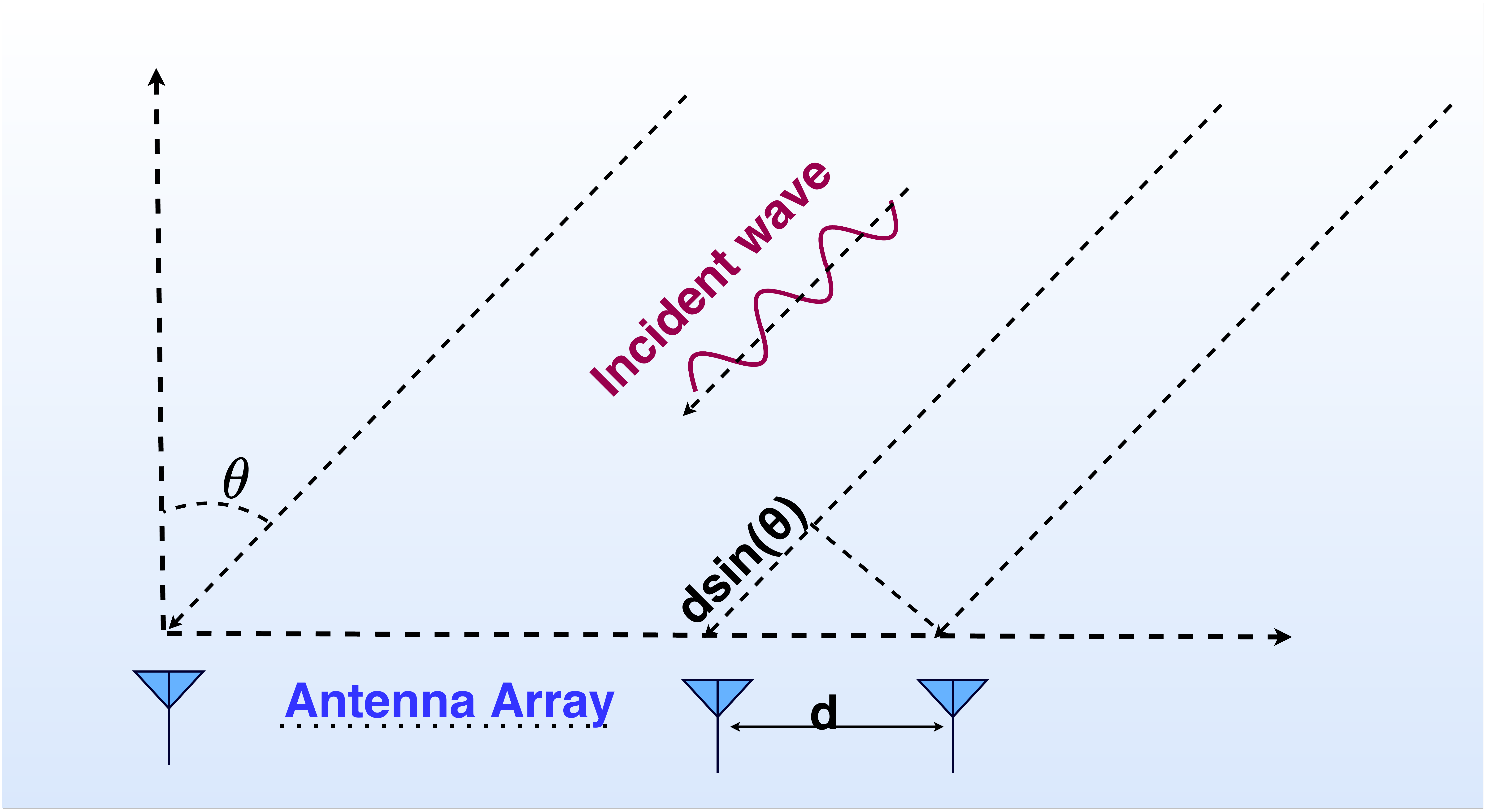}\label{aoafig}}
\hspace{-0.1cm}
\subfigure{\includegraphics[width=4.35cm,height=3.7cm]{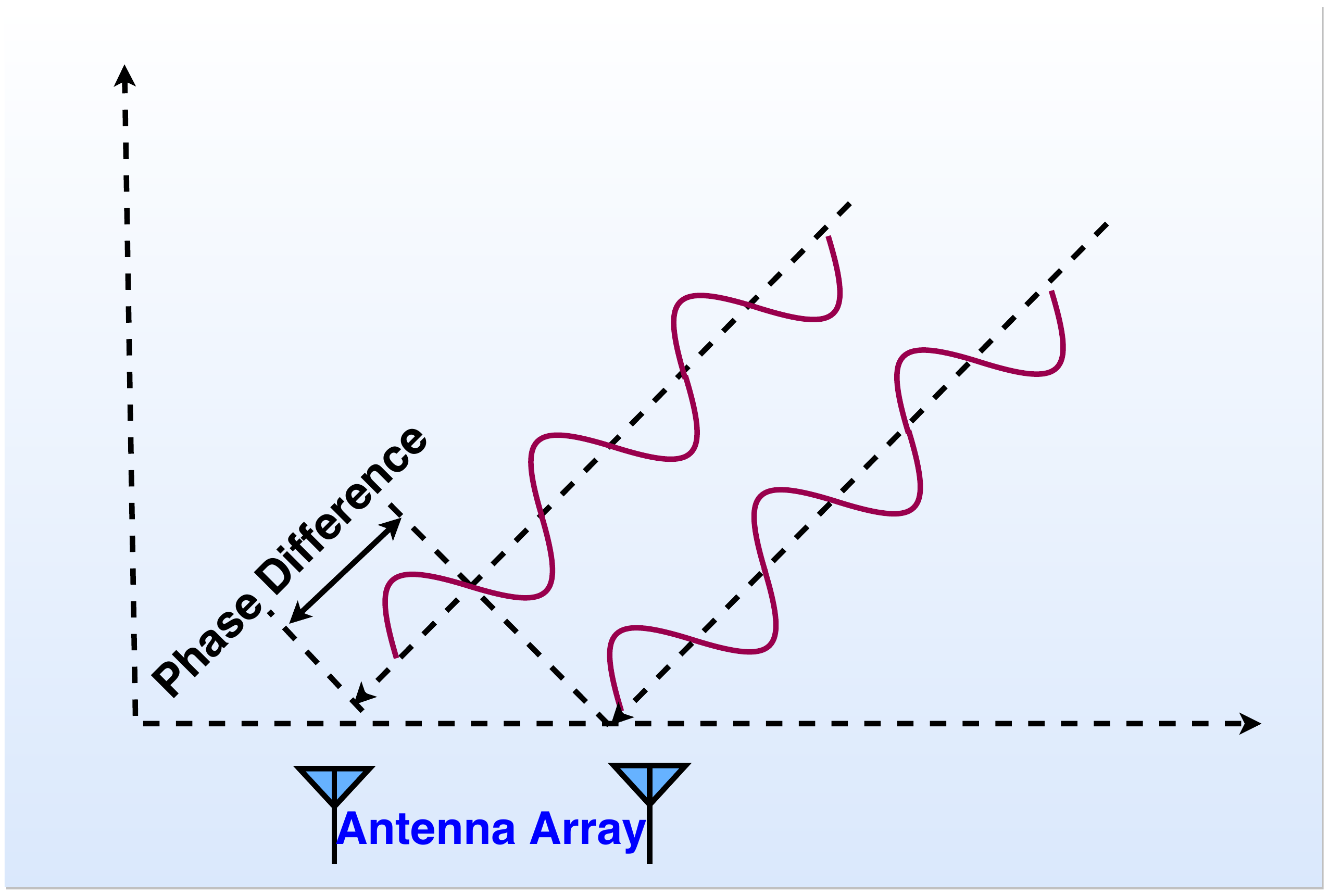}\label{poafig}}

      a) AoA  \hspace{3cm}  b) PoA \\
\caption{ Positioning based on AoA and PoA}
\end{figure}

  \paragraph{   Angle of arrival (AOA)}
In AOA-based localization, the UE's position is determined where pairs of angular direction lines intersect, each circular line formed with the mobile target situated on it. Unlike TOA and TDOA methods, AoA only requires two RNs equipped with antenna arrays. Although more flexible than TOA and RSSI methods, AoA-based localization faces limitations due to obstructions, multipath propagation, and the lack of LoS. OFDM signals have helped alleviate multipath issues. While three measurements suffice for 3D localization and two for 2D, AoA requires extensive hardware and suffers from reduced accuracy proportional to the distance from the RN due to multipath reflection and shadowing \cite{aoa2}. Fig. \ref{aoafig} illustrates AoA's application in approximating the user's position by leveraging signal reception angles.

	\paragraph{Phase of Arrival(PoA)}
Phase of Arrival (PoA) techniques utilize the phase or phase difference of the carrier signal to determine the distance between sender and receiver. The assumption is that signals from anchor nodes or the user device are perfectly sinusoidal with identical frequency and no phase deviation. Methods for estimating the range between the Tx and Rx using PoA include considering a finite transit delay (Di) as a fraction of the signal wavelength. Incident signals reach different antennas within the array with distinct phase disparities, enabling user location derivation. PoA can enhance localization precision when combined with RSSI, ToA, and TDoA, but its reliance on unobstructed line-of-sight poses challenges for accuracy, especially in indoor environments.

 \begin{table*}
 \centering
 \renewcommand{\arraystretch}{1.1}	
		\caption{COMPARISON OF ML-BASED TOOLS FOR IPS. }
		\begin{tabular}{ | m{2.35cm} | m{1.2cm}| m{1.5cm}|  m{2.4cm} |m{1.5cm} |m{1.15cm} |m{0.9cm} |m{1cm} |m{1.1cm} |m{0.8cm}|} 
	
		\rowcolor{lightgray}
			\hline
			 Tool & Complexity & Interpretability & Linearity Status & Computational Cost & Outlier Sensitivity & Memory Intensity & Flexibility &Robustness to Noise & Variance\\ 
    
			\hline
   SVM \cite{svr1,svr2}&Medium&Low&Linear $\&$ non-Linear&Low&High&Low& Low&Low&Low \\
			\hline
   PCA \cite{localization13}&Medium&Low&Linear&High &High&Low&Low&High&Low \\
			\hline
   ANN \cite{visible1,visible2,visible3} &Medium& Medium&Linear $\&$ non-Linear& Medium& High&High&Medium&Low& Medium \\
			\hline
   DNN\cite{localization3,localization6}&High&Medium&Linear $\&$ non-Linear&High&High&Medium&High&Medium& High\\
			\hline
   RNN \cite{visible5,bluetooth6} &High&Low&Linear $\&$ non-Linear&High&High&Low&Low&Medium&High \\
			\hline
   LSTM \cite{localization25}&High&Medium&Linear $\&$ non-Linear&Medium&High&High&Medium&High&High \\
			\hline
   MLP \cite{intro17,uwb5}&Medium&Low&Linear $\&$ non-Linear&Low&Medium&Medium&Medium&High&High \\
			\hline
   CNN \cite{uwb7,uwb6}&High&High&Linear $\&$ non-Linear&High&High&High&High&Medium&High \\
			\hline
   RBF \cite{rfid2,rfid9} &Medium&High&Linear $\&$ non-Linear&Low&Low&Low&Medium&High&High \\
			\hline
   KNN \cite{knn1,knn2} &Low&High&Linear $\&$ non-Linear&High&High&High&Low&Medium& High\\
			\hline
   RF \cite{lid5, bluetooth14} &Medium&High&Linear $\&$ non-Linear&Low&Low&High&High&High&High \\
			\hline
		
	\end{tabular}
		\label{MLcomp}
	\end{table*}

 	\section{PIONEERING SENSING TECHNOLOGIES FOR  INDOOR POSITIONING}

The process of localization involves analyzing data from various sources to determine the location of an object or individual. In the literature, several ML tools have been utilized to analyze and classify the data in different localization approaches. These tools include supervised and unsupervised learning algorithms, deep learning methods, clustering techniques, and regression models. By adopting various ML tools, researchers have been able to achieve high accuracy and efficiency in localization applications, making them a valuable asset in the field of location-based services.  The choice of ML model for an IPS depends on the specific use case and data characteristics. By utilizing these models, IPS can provide accurate and reliable location tracking and data analysis for a variety of indoor environments and applications. A comprehensive comparison of these models, considering factors such as complexity, interpretability, and flexibility, is provided in Table \ref{MLcomp}.
This section explores innovative sensing technologies designed for different IPSs, employing sensors like Wi-Fi, Bluetooth, and magnetic fields for device/object localization. Integrating these sensors with ML models boosts IPS precision, enabling reliable location data for navigation, tracking, and automation. Table. \ref{techcomp} compares these technologies based on coverage area, cost, and energy efficiency.

\begin{table*}[!thp]
 \renewcommand{\arraystretch}{1.1}	
		\caption{INDOOR POSITIONING TECHNOLOGIES. }
		\begin{tabular}{ | m{1.15cm} | m{0.9cm}| m{.8cm} | m{.8cm}| m{1.3cm}|m{0.85cm}|m{4.5cm}| m{4.5cm} |} 
	
		\rowcolor{lightgray}
			
			\hline
				 %\vspace{.2cm}
			 Technology & Coverage & Cost & Energy & Computation& Accuracy & Advantages & Disadvantages \\ 
			\hline

 GPS & 5m & Low&Low &Low &Low& \tabitem Highly ubiquitous \tabitem Wide Area Coverage
 & \tabitem Poor indoor performance \newline
 \tabitem Signal Quality Variability \newline
 \tabitem Signal Interference:\\

        \hline

        IR & 5m& Low & Low  &Low&High& \tabitem Low Signal Interference \newline
        \tabitem Enhanced privacy \newline
        \tabitem Fine-Grained Positioning \newline
        &\tabitem Line-of-Sight Requirement \newline
        \tabitem Sensitivity to Lighting Conditions \newline
        \tabitem Limited Range \newline \\
		\hline
 Magnetic &10m & High &Low &Medium&Low&\tabitem No pre-deployed infrastructure is required\newline 
    \tabitem Magnetic field is everywhere and relatively stable&  \tabitem Complexity increases as the number of sensors increases \newline
    \tabitem Accuracy depends on the variation in magnetic field.\\
\hline
WLAN &45m& Low&High&Medium&High&
 \tabitem Use existing communication networks  \newline
		    \tabitem Majority of devices available nowadays are equipped with WLAN connectivity \newline
            \tabitem LOS is not required  
            % \tabitem Accessibility 
            & \tabitem   Highly influenced by environmental changes.\newline
            \tabitem Security concerns\\

		\hline
  VLC & 60m & Low &  Medium& Medium &High  & \hspace{-0.1cm}\tabitem Supports larger bandwidth \newline 
	\tabitem Secured communication \newline 
	     \tabitem No interference due to EM radiations \newline 
	     \tabitem Easy to install \newline 
	& 
	    \tabitem Interference issues from other  light sources \newline 
	    \tabitem Requires both source and receiver should be in LOS \newline 
	   \tabitem Easily affected by atmospheric absorption, shadowing, and beam dispersion \\
	
		\hline

 RADAR & 70m & Medium&Low&Medium&High&  \tabitem High  Robustness\newline
      \tabitem LOS is not required\newline
      \tabitem Highly  scalable &
      \tabitem  Limited resolution \newline
      \tabitem High chances of signal interference \newline
      \tabitem Complex signal processing\\
\hline

 WSN & 100m&Low&Low&Medium &Medium&
	    \tabitem Suitable for the non-reachable places \newline
	    \tabitem Can accommodate new devices at any time.\newline
	    \tabitem Flexible&  \tabitem Low Speed \newline
	    \tabitem Security issues \newline
	    \tabitem Highly influenced by walls, and microwave\\

		\hline
  RFID&100m & Low& Low & Medium & Medium &\tabitem LOS is not required between RF transmitters and receivers \newline
        \tabitem Penetrate through solid and non-metal objects & 
        
        \tabitem Less compatible with other technologies\newline
        \tabitem The data signal is affected by antenna\newline
        \tabitem Security issues. \\
        	\hline
Zigbee &100m& Low& Medium& Low&Medium
 
 & \tabitem Low power consumption
 \tabitem Offer low latency \newline
 Offers interoperability between devices from different manufacturers.& \tabitem Interference \newline
 \tabitem Limited Data Throughput \newline 
 \tabitem Limited Network Size.\\
	\hline
UWB & 200m &High& Medium &Medium&High& \tabitem Penetrate through walls \newline
		\tabitem No interference with existing RF systems & \tabitem Interference due to metalic and liquid materials\\

		\hline

 Bluetooth &400m&Low&Low&Low&Medium&\tabitem Do not require LOS \newline
		    \tabitem  A lighter standard \newline
		    \tabitem Highly ubiquitous & \tabitem  Relatively expensive\newline
		    \tabitem  High radio interference\\

		% \hline
 
  % Wi-Fi&100m&Low &High& Medium& Low & \tabitem Highly ubiquitous \newline
  % \tabitem Accessibility& \tabitem Security concerns \\

		\hline
 
 Cellular &8km-35km & High & Low & Medium &High&  \tabitem Mobile phones can be used \newline
	    \tabitem  No interference with other devices operating at the same frequency. & \tabitem Low reliability\\

		\hline
 
Acoustic &-& High&Medium&High&High&\tabitem No need of LoS\newline
\tabitem Low Interference\newline
\tabitem Available on Smartphones& \tabitem Sensitive to environmental conditions\\

  		\hline
 Ultrasonic & - &High &Medium &High&High&\tabitem No need of LoS\newline
\tabitem Low Interference\newline
\tabitem Available on Smartphones& \tabitem Sensitive to temperature and pressure\newline
\tabitem No penetration though some solids\\

		\hline
		\end{tabular}
		\label{techcomp}
	\end{table*}

% \vspace{-0.2cm}

		\begin{table*}
   \renewcommand{\arraystretch}{1.1}	
		\caption{VLC $\&$ IR-based indoor positioning systems $\&$ solutions.}
			\begin{tabular}{ | m{3.5cm} | m{3cm}| m{2cm} |  m{1.5cm} |  m{1.5cm} | m{4cm} | } 
			\rowcolor{lightgray}
			
			\hline
				% \vspace{.2cm}
			 Technology &  Positioning  algorithm  & Complexity & Scalability  & Cost & Paper \\ 
			\hline

			% \vspace{.4cm}
			 Visible Light & ANN & Medium & Low & Low &  \cite{visible1, visible2,visible3,  visible4,visible7,localization35}\\
			\hline

			 % \vspace{.2cm}
				 Visible Light & Regression NN  & Medium & Medium & Low & \cite{visible5} \\
     
			\hline
			
			% \vspace{.2cm}
			
		 Visible Light & DNN &  High & Low & Low &\cite{visible6} \\
			\hline

% \vspace{.2cm}

             PIR Sensor &DT&Low&Low&Low& \cite{IR6}\\ \hline
			% \vspace{.2cm}

         PIR Sensor &XGB, SVM, LR, RF&Medium&Low&Low& \cite{IR7}\\ \hline
			% \vspace{.2cm}

   % \vspace{.2cm}

       PIR Sensor&ANN &Medium&High&Low& \cite{IR9}  \\ \hline

            % \vspace{.2cm}
    PIR Sensor &CNN-BiLSTM& High&Low &Low& \cite{IR3}\\
            \hline

	% \vspace{.2cm}
              PIR Sensor &Active transfer learning  &High&Low&Low& \cite{IR8} \\ \hline
		
			% \vspace{.2cm}
PIR Sensor&CNN& High&Medium &Low& \cite{IR2}\\
            \hline

   %          IR &&&&& \cite{IR3}\\ \hline
			% \vspace{.2cm}

             PIR Sensor& DCNN, LSTM, RNN &High&High&Low& \cite{IR4,IR5}\\ \hline
			% \vspace{.2cm}

            IR Sensor&SVM &Low&Low&Low& \cite{IR10,IR11}  \\ \hline
            % \vspace{.2cm}
    IR Thermal Sensor&DCNN&Low&Low&Low& \cite{IR15}  \\ \hline

            % \vspace{.2cm}

            IR Thermal Sensor&RF, GNB, kNN, SVM &Medium&Low&Low& \cite{IR12}  \\ \hline

            % \vspace{.2cm}

            IR Thermal Sensor&SVM,FFNN &Medium&High&Low& \cite{IR13}  \\ \hline

            % \vspace{.2cm}

            IR Thermal Sensor&DNN &High&Low&Low& \cite{IR14}  \\ \hline

		\end{tabular}
		\label{VLCIR}
	\end{table*}

	% \vspace{-0.2cm}

	\subsection{Visible Light Communication (VLC)-based}

	Visible light communication (VLC) using LEDs has garnered attention due to its energy efficiency and cost-effectiveness, especially for indoor illumination, and its resilience to electromagnetic interference. Consequently, research efforts have intensified to improve indoor visible light positioning (VLP) systems. Photo-diodes (PDs) are preferred over image sensors for VLC detection due to their affordability and simplicity, particularly with the RSS algorithm, which is more reliable in larger scenarios \cite{localization35}. However, RSS-based VLP suffers from positioning errors, prompting researchers to explore ML-based methods to enhance accuracy.
 
Various ANN-based visible light indoor positioning systems are proposed in \cite{visible1}, \cite{visible2} and \cite{visible3}. In \cite{visible1}, an LED bulb is considered as the transmitter, and a photo diode is considered as the receiver. The communication channel is modeled as a visible light channel influenced by multipath effects which is removed by the application of ANN.   The system used the Combined Deterministic and Monte Carlo (CDMMC) method for localization.  In \cite{visible2}, the system used a back-propagation ANN with optical camera communications where LED lights are clustered into blocks and these coordinates are encoded using under sampled modulation scheme. The location of the camera { receiver) is localized using the estimated coordinates of the blocks. 
Two indoor positioning methods using VLS -based on ANN are proposed by Shencheng  $ et ~ al.$ \cite{visible3}. In this approach, the first method creates 4 networks that are used for training and estimating the coordinates of the target location. The second method creates only one network with 4 input layers. The system divides the entire indoor space into small blocks. The receiver captures the signal sent by each LED with ID information with which the system localizes the receiver. The transmitted optical signal is modulated using asymmetric clipped optical orthogonal frequency division multiplexing (ACO-OFDM) technology. These 2 methods are robust and achieved a mean position error of 3.29cm and 2.78cm, respectively.
			
The utilization of VLC combined with a genetic algorithm in indoor localization, as presented in \cite{localization35}, involves calculating 3D coordinates within the optical wireless environment using multipath reflections. A modified genetic algorithm facilitates global optimization for 3D position estimation without prior knowledge of device height or orientation. ANN processes first-order reflections, achieving a low average localization error of 1.02 cm. 3D visible light positioning (3DVLP) proposed in \cite{visible5} excels in accuracy and consistency. It employs a regression neural network to estimate real-time target location. Image sensors capture AoA information from transmitting LED lights for network input. The offline phase preprocesses the data extracted from reference points, which trains the network using Adam optimizer. The trained network in the online phase detects target locations with a mean error of 1.1 cm.

Sayed $ et ~ al.$ introduced two VLC-based methods to locate a moving user in \cite{visible4}. VLC transmitters convey location info through visible light, which a photodetector carried by the target receives. Trilateration and neural network approaches are used for instantaneous prediction. For LOS, the system's max positioning error is 2.9 cm; for NLoS, it's 8.1 cm, with the neural network outperforming trilateration. In \cite{visible6}, a Position Estimation Deep Neural Network (PE-DNN) aided VLC system is introduced to address complexity and compatibility issues. X. Lin's approach employs a DNN for processing data, enabling 2D location estimation with just one LED transmitter. Achieving centimeter-level accuracy, the system attains a minimum positioning error of 4.18 cm. In \cite{visible7}, a trained neural network mitigates the impact of indoor diffuse channels in VLC positioning. Utilizing RSS data and backpropagation, the algorithm achieves an average positioning error of 6.59 cm.

VLC-based IPSs encounter challenges due to signal blockage, lighting conditions, and LoS constraints. Complex algorithms for angle estimation and decoding, specialized hardware, and integration with lighting systems affect cost and scalability. Addressing these is essential for improving VLC-based IPS accuracy and feasibility. A comparison of VLC-based IPSs considering factors like complexity, scalability, and cost is given in Table.  \ref{VLCIR}.

% \vspace{-0.2cm}
   
			\subsection{Infrared (IR)-based}
.
In Infrared (IR) technology for IPS, electromagnetic radiation beyond the visible light spectrum is utilized. An IR system comprises an emitter diode emitting bursts of non-visible light as infrared signals. A receiving photo diode captures these light pulses for processing. IR positioning functions in active or passive modes, with system reliability contingent on optical signal properties such as directivity and obstacle response, encompassing reflectivity and scattering \cite{survnew5}.  In IR-based IPSs, the LOS requirement poses challenges due to occluded no-detection zones from the transmitter or sensor. This technology is frequently applied in robotics, smart homes, and indoor navigation systems to furnish real-time location information and enhance user experiences indoors.

Numerous research works have been suggested for IPS employing CNN as the analytical technique \cite{IR2, IR3, IR4, IR5}. Recent developments target reduced sensor density through analyzing analog output \cite{IR3}. Using PIR and domain expertise, \cite{IR3} introduces a CNN-based multi-person localization system with modules for person count and location determination. It attains 76$\%$ density reduction while retaining accuracy, employing two-stage networks for signal separation, detection, extraction, and localization in complex scenarios. Comparing classical and deep learning algorithms for analog PIR-based human movement detection, \cite{IR2} demonstrates CNN's superior real-time detection and accuracy, even with limited data, while assessing aspects like scalability and real-time performance. Leveraging the PIR sensor's analog output's reflection of temperature changes, \cite{IR4} proposes a CNN-LSTM model capturing temporal dependencies for accurate location estimation, validated in intricate scenarios. Similarly, \cite{IR5} presents a PIR-based system and dataset, applying CNN, RNN, and CNN-RNN for device-free localization feasibility. Dataset conversion yields a remarkable 0.25m distance error, spotlighting PIR's accurate localization potential \cite{IR5}.

An alternative method within PIR-based Indoor Positioning Systems involves utilizing thermal imaging to improve localization accuracy\cite{IR8, IR6, IR7}.  In \cite{IR8}, a thermal imaging-based localization system proves useful in emergencies, utilizing active transfer learning for accurate location determination in dark conditions. Meanwhile, \cite{IR6} presents a mobile robot equipped with an IR thermal imaging camera for enhanced localization precision through overlapping fields of view and a Decision Tree classifier, achieving 96$\%$ accuracy in controlled settings but facing constraints (83.3$\%$) in unconstrained scenarios. In passive human localization, \cite{IR7} employs infrared thermal imaging cameras to detect emitted IR radiation, employing a machine learning-based approach utilizing human body temperature for precise person-to-camera distance and localization estimation. Factors like head position, size, and temperature statistics enable sub-meter accuracy, comparing ML tools such as XGB, RF, SVM, and LR for person-to-camera distance estimation. 

Another application of IR thermal imaging sensors is to use them for precise tracking of patients and caregivers. This can be facilitated through discreet approaches such as utilizing low-resolution infrared sensors to protect privacy\cite{IR4, IR10, IR11}. The objective of the research in \cite{IR10} is to develop a cost-effective indoor localization system that prioritizes privacy, achieved by employing a network of infrared sensors with the help of an SVM classifier. Similarly, \cite{IR11} also proposed a system to achieve motion detection and group proximity modeling with an 8x8 infrared sensor array using SVM. Each pixel provides temperature readings, forming scenes with different human group configurations and walking directions. Motion direction is inferred using cross-correlation analysis, while the SVM classifier estimates the number of human subjects in each scene. Similarly, the proposed approach in \cite{IR14} utilizes a low-resolution infrared array sensor and Deep Learning to achieve up to 97$\%$ accuracy in detecting up to 3 people's presence and 100$\%$ accuracy in determining their absence. The study in \cite{IR15} focuses on unobtrusive human posture recognition for health monitoring, especially for the elderly with the help of an infrared sensor-based wireless network and DCNN.  Another low-resolution thermal imaging sensor for real-time localization in smart buildings is proposed in \cite{IR12}. Different sensors with varying resolutions are analyzed for their localization performance. The paper presents a unified processing algorithm pipeline and proposes various algorithms for data preprocessing, feature extraction, and localization. A privacy-preserving indoor localization system using IR thermal imaging sensors and machine learning to collect and analyze toilet usage data in an office is proposed in \cite{IR13}. Evaluating occupancy patterns contributes to the evidence-based design of sanitary spaces and adds to scholarship on indoor localization methods. The comparison of these LiDAR-based systems is given in Table. \ref{sound}.

IPSs utilizing IR face significant challenges. LoS dependency affects accuracy due to obstacles, while interference from ambient light sources disrupts signal integrity. Multipath effects introduce complexities in signal processing. Signal attenuation caused by distance, absorption, and scattering complicates distance estimation. Limited signal range necessitates dense sensor deployment, raising infrastructure costs.  Ensuring privacy amid data capture and infrastructure maintenance adds complexity. Addressing these challenges requires holistic solutions integrating technology, algorithms, and system design. The table in \ref{VLCIR} provides a comparison of IR-based IPSs considering factors like complexity, scalability, and cost. 

   \subsection{LiDAR-based}
LiDAR-based IPS utilizes Light Detection and Ranging (LiDAR) technology to enable precise and accurate localization of objects or devices within indoor environments. LiDAR technology emits laser pulses and measures the time it takes for the pulses to return after bouncing off surfaces, creating detailed 3D maps of indoor spaces. These maps can then be used for various applications, including indoor navigation, tracking, and automation.  The study proposed in \cite{lid1} fuses CNN-predicted depth maps with direct monocular Simultaneous Localization and Mapping (SLAM) depth measurements. This fusion approach improves reconstruction quality in challenging areas. The depth prediction also aids in absolute scale estimation, addressing a limitation of monocular SLAM. Additionally, a framework is introduced to fuse semantic labels with dense SLAM, enhancing scene reconstruction from a single view.  Similar to this, a DL  architecture for real-time dense mapping and tracking is proposed in \cite{lid2}. Synthetic viewpoints aid incremental tracking, simplifying the learning process. The proposed system accumulates information in a cost volume for accurate depth estimations by combining depth measurements and image-based priors.
\cite{lid3, lid4} introduced a system that seamlessly integrates fingerprinting-based indoor localization with minimal data collection effort. LiDARs are leveraged to tag Wi-Fi scans during regular user movement, reducing human intervention. These systems can efficiently build the fingerprint database using even a single LiDAR, allowing reuse across buildings and minimizing deployment costs. 
An alternative strategy for incorporating LiDARs into IPSs involves the utilization of compact wearable micro LiDARs for localization \cite{ lid5}. Such a system is introduced in \cite{lid5} for accurate human identification in real-time life-logging applications. The system utilizes 3D point cloud data extracted from the LiDAR sensor, removing noise and background through Spatio-temporal density clustering and fisher vector representations. A random forest classifier is trained on the extracted features, achieving 99.9$\%$ accurate subject identification. 

LiDAR-based IPSs face distinct challenges. Firstly, LiDAR sensors require LoS visibility, hindering accurate positioning in obstructed environments. Secondly, dense indoor spaces can lead to multiple reflections and scattering, causing signal distortions. Thirdly, the cost and complexity of LiDAR sensors can limit their widespread adoption. Moreover, real-time processing of LiDAR data demands powerful computational resources. Calibrating LiDAR sensors for accurate measurements and addressing issues related to sensor drift and environmental changes also pose challenges. These factors collectively impact the accuracy and feasibility of these systems.
% \vspace{-0.2cm}
\begin{table*}
 \renewcommand{\arraystretch}{1.1}	
		\caption{LiDAR and Sound-based indoor positioning systems $\&$ solutions.}
			\begin{tabular}{ | m{3.5cm} | m{3cm}| m{2cm} |  m{2cm} |  m{2cm} | m{3cm} | } 
			\rowcolor{lightgray}

			\hline

			% \vspace{.2cm}
			 Technology &  Positioning  algorithm  & Complexity & Scalability  & Cost & Paper \\ 
			\hline

			LiDAR Sensor & RF & Low & Medium & High &  \cite{lid5} \\ \hline

 % \vspace{.2cm}

  LiDAR Sensor  $\&$ RSSI &  LSTM,SVM & Medium & Medium & High &  \cite{lid3,lid4} \\
			\hline
					% \vspace{.2cm}
			
		  LiDAR Sensor & DL & High & Low & High &  \cite{lid2} \\
			\hline
		
				% \vspace{.2cm}
	LiDAR Sensor & CNN & High & Medium & High &  \cite{lid1} \\
			\hline

	% \vspace{.2cm}
		  Acoustic & DNN & High & Medium & Medium & \cite{sound3,sound4}\\
			\hline	
			% \vspace{.2cm}
			  Acoustic & PNN & High & Medium & High & \cite{sound1} \\
			\hline

			% \vspace{.2cm}
			  Acoustic & DNN & High & Medium & High & \cite{sound2} \\
			\hline

			% \vspace{.2cm}
			 Acoustic & CRNN &  High & Medium & High & \cite{sound5}\\
			\hline
			
			% \vspace{.2cm}
			 Acoustic & CNN &  High & Medium & High & \cite{sound6}\\
			\hline

	% \vspace{.2cm}
	% 		 Ultrasonic & - &  High & Medium & High & \cite{ultra8}\\
	% 		\hline

% \vspace{.2cm}
% 			 Ultrasonic & CNN &  High & Medium & High & \cite{proximity1}\\
% 			\hline

	% \vspace{.2cm}
	% 		 Ultrasonic & CNN &  High & Medium & High & \cite{ultra9}\\
	% 		\hline
        % \vspace{.2cm}
			 Ultrasonic & SVM &  High & Medium & High & \cite{ultra1}\\
			\hline
	% \vspace{.2cm}
			 Ultrasonic & SVM &  High & Medium & High & \cite{ultra10}\\
			\hline
	% \vspace{.2cm}
			Ultrasonic & SVM &  High & Medium & High & \cite{ultra11}\\
			\hline

            % \vspace{-0.2cm}

		\end{tabular}
  \label{sound}
	\end{table*}

% \vspace{-0.2cm}
 \subsection{Sound-based}

    Sound Source Localization (SSL) is a vital process that determines the origin of sound in an environment, crucial for applications like acoustic surveillance, robot localization, virtual reality, and human-computer interaction. SSL methods utilize microphone arrays strategically placed to capture sound signals, analyzing time delay, phase differences, or intensity variations to estimate sound source direction. Techniques like TDOA, IDOA, and PDOA are employed, often combined with signal processing algorithms, to achieve accurate localization in diverse scenarios and noise conditions. SSL's significance lies in enhancing situational awareness, aiding robotic navigation, and enabling immersive audio experiences in various real-world settings.
    \subsubsection{Acoustic-based}
			
SSL finds utility in various domains like human-robot interaction, teleconferencing, and automatic speech recognition \cite{sound1}. For instance, SSL enables robots to locate patients efficiently. Nonetheless, SSL poses challenges due to multipath effects, high reverberation, and low signal-to-noise ratio (SNR). To tackle these issues, \cite{sound2} introduced an SSL algorithm employing a probabilistic neural network (PNN) called a generalized cross-correlation classification algorithm (GCA). GCA transforms the SSL problem into a likelihood-based nonlinear classification using PNN. To enhance accuracy and Direction of Arrival (DOA) estimation, they utilized the weighted location decision method (WLDM). Compared to existing methods, their approach yields azimuth angle estimation errors averaging 4.6$^{\circ}$ and elevation angle estimation errors averaging 3.1$^{\circ}$. Additionally, a DNN-based SSL system for indoor environments is proposed in \cite{sound1}. It efficiently locates sound sources in various positions. Microphones capture sound signals, and spatial features are extracted and represented as likelihood surfaces. An encoder compresses input data, while decoders estimate source locations.

                Many more works are proposed for SSL using various ML tools \cite{sound3,sound4}. In \cite{sound3}, an unsupervised adapting DNN is used for classification.  The influence of unknown reverberant environments is highly challenging in SSL.  However, they solved this problem by using unsupervised adaption of parameters of DNNs to the collected sound signals.
                \cite{sound4} studied the potential of DL-based time-frequency (T-F) masking in SSL. It enhances DOA estimation, especially in reverberant conditions. Speech-dominant T-F units containing relatively clean phases for DOA estimation are identified by DNN. 
                Another SSL technique using convolutional recurrent
                neural network for joint sound event localization is proposed in \cite{sound5}. This system is capable of detecting multiple overlapping sound events in 3D space. The network feeds the sequences of consecutive spectrogram time frames as inputs. It maps to two outputs parallelly : sound event detection and localization in 3D space. Here, detection is considered as a multi-label classification problem and localization is a multi-output regression task. This system uses the phase and magnitude component of the spectrogram calculated on each audio channel as the feature separately.
             
                One more paper proposed a similar approach in SSL  using CNN \cite{sound6}. In this paper, a supervised learning method for estimating the
                DOA of multiple speakers is proposed. Hence, the estimation of multi-speaker DOA is considered a multi-class multi-label
                classification problem. Here, the phase component of the short-time Fourier transform (STFT) coefficients of the received signals from the microphone are fed into the CNN, and the feature estimation of DOA is carried out during training.
                	\subsubsection{Ultrasound-based}

				Systems using ultrasound signals for location estimation are inspired by the navigation techniques of bats at night. From this idea \cite{ultra1} proposed an Active Bat positioning system to provide 3D localization for the tags.  Tags are small tracked devices carried by people and of dimension 7.5cm $x$ 3.5cm $x$ 1.5 cm.  Active Bat system adopts ultrasonic technology and a triangulation approach to detect the location of the tag-carrying person. 

    The paper \cite{ultra10} introduces the Acoustic Location Processing System (ALPS), a platform that enhances BLE transmitters with ultrasound for improved ranging accuracy in location-aware applications. By placing three or more beacons and performing a calibration sequence, precise beacon locations and room geometry are computed. ALPS utilizes time-synchronized ultrasonic transmitters and achieves an estimated 16.1cm 3D beacon location error and 19.8cm room measurements error. It can identify NLoS signals with over 80$\%$ accuracy and track a user's location within 100cm. SVM is employed to filter out NLoS signals during user localization after installation.
Another work \cite{ultra11} focuses on sound-based localization technologies for indoor positioning, compatible with smartphones and cost-effective. However, the NLoS phenomenon poses challenges. The study proposes an efficient approach using SVM with a radial-based function kernel to identify NLoS components by characterizing acoustic channels. Nine novel features are extracted, achieving an overall 98.9$\%$ classification accuracy with a large dataset of over 10 thousand measurements. The SVM with RBF kernel outperforms traditional classifiers in acoustic NLoS identification.
The comparison of these works is given in Table. \ref{sound}.
            
IPSs that utilize acoustic and ultrasonic signals face significant challenges. Sound waves are easily affected by reflections, absorptions, and diffractions, causing signal distortions and inaccuracies in position estimation. Background noise further hampers signal quality, affecting accuracy. Precise synchronization and timing are essential for distance calculation while varying propagation due to environmental factors adds complexity. These challenges collectively influence the reliability and performance of sound-based IPSs.

 \subsection{RADAR-based}

               RADAR (Radio Detection And Ranging) systems, utilizing electromagnetic waves, locate objects by transmitting high-frequency radio waves reflected off encountered objects, like airplanes or ships. By analyzing the reflected wave's characteristics, RADAR determines object range, direction, and speed. In Indoor Positioning Systems (IPS), RADARs have diverse applications, including human detection, tracking, and localization. IPS benefits from RADARs' unique traits, like their ability to penetrate obstacles and high measurement accuracy for distances, velocities, and angles. In \cite{Rad1}, Bahl et al. presented a RADAR-based IPS similar to k-Nearest Neighbours (kNN), offering improved accuracy using signal propagation models. There are different RADAR types, like pulsed, continuous-wave, and FMCW RADARs, each have distinct advantages and are suitable for specific applications. We emphasize FMCW RADAR due to its widespread use in localization tasks.
                
% \vspace{-0.2cm}
\begin{table*}
 \renewcommand{\arraystretch}{1.1}	
		\caption{RADAR-based indoor positioning systems $\&$ solutions.}
		\begin{tabular}{ | m{3.5cm} | m{1.5cm} | m{2.5cm}| m{2cm} |  m{2cm} |  m{2cm} | m{1.5cm} | } 
			\rowcolor{lightgray}
			\hline

				% \vspace{.2cm}
			
			% \vspace{.2cm}
				 Technology & Frequency & Positioning  algorithm  & Complexity & Scalability  & Cost & Paper \\ 
			\hline

        % \vspace{.2cm}
            IWR6843 Radar &60GHz& DNN   & Medium  & Low & Medium& \cite{fmcw2}\\

            \hline
% \vspace{.2cm}
             IWR6843 Radar &60GHz& DBSCAN  & Medium & Low & Medium & \cite{fmcw14}\\
             \hline
            
            % \vspace{.2cm}
             AWR1243 Radar & 79GHz& SVM & Medium & Low & Medium & \cite{fmcw18}\\
             \hline
% \vspace{.2cm}
            60 GHz sensing Radar &60 GHz  & LSTM & Medium & High & High & \cite{fmcw9}\\
        \hline
             
% \vspace{.2cm}
            AWR1443 Radar & 77-81 GHz & CNN & High & Low & Low & \cite{fmcw10}\\
        \hline
        % \vspace{.2cm}
  IWR1443BOOST Radar &77GHz&GaitCube & High &Low &Medium& \cite{fmcw8}\\

            \hline
% \vspace{.2cm}
           AWR1642 Radar &77GHz& CNN & High & Low & Medium& \cite{fmcw5}\\
            \hline
      
             % \vspace{.2cm}
             IWR1443 Radar &76-81 GHz& DBSCAN & High & Low & Medium & \cite{fmcw16}\\
             \hline

             %\vspace{.2cm}
             AWR1843BOOST Radar &77GHz& RNN AutoEncoder & High & Low & Medium & \cite{fmcw17}\\
             \hline
% \vspace{.2cm}
           AWR1443  Radar &76-81 GHz& DBSCAN &High & Low & High& \cite{fmcw12}\\
        \hline

% \vspace{.2cm}
             IWR1642 Radar &77GHz& CNN & High & Low & High & \cite{fmcw13,fmcw15}\\
             \hline

        %\vspace{.2cm}
            INRAS RadarLog devic & 77GHz& DBSCAN, DCNN & High & High & Medium  & \cite{fmcw11}\\
        \hline
             % \vspace{.2cm}
           AWR1642BOOST Radar & 77GHz & DCNN&High&High&Medium& \cite{fmcw6}\\

            \hline

		\end{tabular}
  \label{radar}
	\end{table*}
 
\subsubsection{Frequency Modulated Continuous Wave (FMCW)-based}

Frequency Modulated Continuous Wave (FMCW) radar uses frequency modulation to determine the range of targets, emitting a continuous wave with a linearly varying frequency. The received signal is mixed with a reference signal, producing an Intermediate Frequency (IF) signal which is proportional to the range of the target. By measuring the frequency difference between the transmitted and received signals, FMCW radar can determine the range of the target \cite{fmcw1}. 
FMCW radar can also provide information on the target's velocity and AoA, making it a popular choice for indoor positioning systems.  
Modern radars often use FMCW modulation along with Millimeter wave (mmWave) technology to provide accurate range and velocity information. mmWave radars are sensors with high resolution and low power, capable of detecting targets and providing range information regardless of weather conditions or location.
Various systems have adopted mmWave-based point cloud data processing for object detection and localization. FMCW radars can have a beamwidth and field of view (FoV) adaptable up to 120$^o$, with a maximum range of 300m. Multiple radars can be cascaded to achieve a wider FoV. Time-frequency analysis is also used to investigate further properties of the data, increasing the complexity of signal processing. Therefore,  ML-based approaches have been adopted to process the radar's data.
% {\color{red}references added,  fill up }
When mapping the radar signal onto the environment, the resulting data points are densely packed in terms of range, angle, and velocity. Within radar sensing, a notable algorithm for this purpose is the density-based spatial clustering of applications with noise (DBSCAN) \cite{fmcw3}. DBSCAN employs a density-based approach to group points based on their proximity in regions of high density while disregarding isolated points. The algorithm starts by selecting an initial point as a core point due to its dense neighborhood. Other points within the neighborhood of the core point, within a specified radius, are considered reachable points. DBSCAN evaluates the neighborhood density of each reachable point and designates new core points from those within dense neighborhoods. By connecting these dense neighborhoods, clusters with diverse shapes are formed, consisting of closely connected points. 
This iterative process continues until no additional reachable points exceed the minimum density requirement.

FMCW mmWave radar has emerged as a prominent technology for human detection and tracking, capitalizing on its capabilities like high resolution and material penetration. Notably, \cite{fmcw8} introduced an inconspicuous human recognition system employing a novel 3D gait data cube representation, derived from FMCW mmWave radar micro-Doppler and micro-range signatures. Signal processing techniques, known as GaitCube, facilitated automatic human walking detection and segmentation, extracting gait data cubes for analysis. Other studies like \cite{fmcw9,fmcw10,fmcw11,fmcw12} proposed diverse human detection, identification, and tracking systems utilizing FMCW radars. mmSense\cite{fmcw9} harnessed mmWave's distinct capabilities, employing LSTM-based classification to detect and localize multiple individuals without devices. milliMap \cite{fmcw10} offered an indoor mapping system for low-visibility settings, integrating lidar during training for noise mitigation, and leveraging spectral responses of mmWave reflections for object identification. Meanwhile, \cite{fmcw11} introduced a system employing backscattered mm-wave signals for simultaneous tracking and recognition of individuals navigating indoor spaces, employing KFs and DNNs for multiple-person scenarios. Finally, \cite{fmcw12} demonstrated a real-time tracking system using two mmWave radars with DBSCAN to efficiently detect and track individuals. These approaches underscore the potential of FMCW mmWave radar for diverse indoor applications.

Human motion behavior detection using FMCW radars is a burgeoning research domain within sensing technology. These radars analyze micro-Doppler signatures generated by human body movements to detect various activities, postures, and individuals, and even monitor vital signs. The paper \cite{fmcw5} presents a real-time human behavior detection system using AWR1642 radar, employing micro-Doppler information captured by the radar and a CNN for classification. Similarly, \cite{fmcw6} introduces a patient behavior detection system combining mmWave radar and deep CNN for real-time recognition of patients' behaviors. The radar tracks and monitors patients, analyzing their scattering point cloud's Doppler pattern using a deep CNN for accurate behavior recognition. Human motion identification is addressed in \cite{fmcw7,fmcw2}, \cite{fmcw7} employs parallel signal processing methods involving statistical analysis and deep learning-based classification of radar spectrograms. And \cite{fmcw2} utilizes a complex-weighted learnable pre-processing module named CubeLearn to directly extract features from raw radar signals, subsequently employing an end-to-end DNN for consistent classification accuracy. These studies collectively demonstrate FMCW radars' potential for sophisticated human behavior analysis and identification.
The application of mmWave radar in sensing extends to human pose estimation, a crucial aspect of human-computer interaction. This involves accurately recognizing body parts like shoulders, wrists, and ankles, enabling the creation of dynamic human body skeletons. This technique finds use in intelligent surveillance, gaming, activity analysis, and smart homes. Pose estimation approaches are presented in \cite{fmcw13,fmcw14,fmcw15}. In \cite{fmcw13}, a system uses data from two radar sensors to generate heatmaps, which a CNN then converts into human poses. Training the network involves coordinated heatmaps from radar and camera inputs. The approach in \cite{fmcw14} categorizes postures (standing, sitting, lying) for an individual indoors using DBSCAN. Likewise, \cite{fmcw15} introduces mm-Pose, a real-time system detecting over 15 skeletal joints via mmWave radar reflection signals. It employs a CNN with a forked architecture to estimate the 3D positions of skeletal joints. These approaches illustrate mmWave radar's potential in accurate human pose estimation for diverse applications.

Another emerging application of FMCW mmWave radar is human activity recognition (HAR). By analyzing the reflected radar signals, such as Doppler or micro-Doppler signatures, it is possible to detect and recognize different activities performed by individuals, such as walking, running, sitting, or gesturing.   \cite{fmcw16} has explored the utilization of FMCW mmWave radars in combination with ML techniques (DBSCAN) as part of their human activity recognition (HAR) approaches. The integration of FMCW radars into health monitoring systems represents a specific subset of HAR. These systems are designed to monitor various aspects of human activity, including fall detection \cite{fmcw17} and identifying sleep positions \cite{fmcw18}, among other functionalities.  All these studies are compared in Table. \ref{radar}

Usually, radar signals face multipath reflections, resulting in imprecise distance measurements and positioning. Further, signal interference from various electronic devices and sources can compromise accuracy. Sophisticated signal processing is essential to discern and analyze distinct reflections. Additionally, the presence of dynamic objects or individuals in the surroundings can influence performance. These combined challenges influence the dependability and resilience of RADAR-based IPSs.

	\begin{table*}
  \renewcommand{\arraystretch}{1.1}	
		\caption{Magnetic Sensor-based indoor positioning systems $\&$ solutions.}
			\begin{tabular}{ | m{3.5cm} | m{3cm}| m{2cm} |  m{2cm} |  m{2cm} | m{3cm} | } 
			\rowcolor{lightgray}
			
			\hline
			% \vspace{.2cm}
			 Technology &  Positioning  algorithm  & Complexity & Scalability  & Cost & Paper \\ 
			\hline

         % \vspace{.2cm}
			 Geomagnetic Sensor & TCN, LSTM &  Low & High & Low &  \cite{magneticnew1} \\
			\hline

			% \vspace{.2cm}
			 Geomagnetic Sensor & RNN &  Medium & Low & Medium &  \cite{magnetic6} \\
			\hline

            % \vspace{.2cm}
			 Geomagnetic Sensor & DNN &  High & Medium & Low &  \cite{magneticnew3,magneticnew4,magneticnew9} \\
			\hline

% \vspace{.2cm}
			 Geomagnetic Sensor & ELM  &  High &Medium& Low &  \cite{magneticnew8} \\
			\hline
         % \vspace{.2cm}
	 Geomagnetic Sensor & DNN  & High & Medium & High &  \cite{magnetic1} \\
			\hline
          % \vspace{.2cm}

		  Magnetic Sensor & ANN  & Medium & Low & Low & \cite{magnetic2} \\
			\hline
			
            % \vspace{.2cm}
			Magnetic Sensor &  XGBoost &  Medium & Low & Low &  \cite{magneticnew2} \\
			\hline
            % \vspace{.2cm}
			 Magnetic Sensor &  ANN &  Medium  & Medium & Low &  \cite{magneticnew5} \\
			\hline

   		% \vspace{.2cm}
			Magnetic $\&$ light sensors  & Deep LSTM & Medium & High & Medium & \cite{localization25}   \\
			\hline

% \vspace{.2cm}
			 Magnetic Sensor & DCNN, mkNN &  High& High & Low &  \cite{magneticnew6} \\
			\hline

% \vspace{.2cm}
			 Magnetic Sensor & CNN &  High & High & Low &  \cite{magneticnew7} \\
			\hline

\end{tabular}
		\label{magnetic}
	\end{table*}
   % \vspace{-0.2cm}
			\subsection{Magnetic Sensors-based}
			
			\par 
   Magnetic sensors are integral to IPS, detecting Earth's magnetic field changes indoors to ascertain device position and orientation. Variations caused by objects or architecture create distinct patterns for localization. Geomagnetic fingerprinting involves comparing real-time readings with a magnetic database. Offering stability and simplicity, magnetic sensors are preferred in indoor settings with unreliable RF signals. Machine learning, like recurrent neural networks, enhances accuracy by analyzing magnetic data sequences, strengthening IPS performance. Positioning and tracking systems using magnetic signals offer high accuracy even in NLoS scenarios \cite{magnetic1}.  
   
    Existing magnetic positioning methods encounter challenges in wide-area accuracy due to magnetic data ambiguity, often requiring multiple sensors. Addressing this, \cite{magnetic1} proposes an indoor system utilizing distorted geomagnetic fields. Features from magnetic sequences are extracted, and fed into a trained neural network alongside a magnetic map, yielding 2D locations with 80$\%$ accuracy. Similarly, \cite{magnetic2} employs an ANN with 5 hidden layer neurons for nonlinear input-output mapping, enhancing self-sensing active magnetic bearing systems.

        Existing RF-based indoor positioning algorithms are unsuitable for large-scale areas like airports due to proportional positioning errors. A geomagnetic sensor-based method proposed in \cite{magnetic6} uses stable geomagnetic data for indoor localization. Object movement affects geomagnetic signals, and RNN models track signal variations for target position detection. RNNs recognize time-varying sensor data sequences. Training and testing occur using the indoor space's magnetic field map, tuning hyperparameters. TensorFlow and CUDA Toolkit are used, achieving positioning errors of 0.51m and 1.04m for medium and large spaces, respectively.
        Another magnetic localization method is proposed using a multi-scale temporal convolutional network (TCN) and LSTM in \cite{magneticnew1}. Time-series preprocessing enhances geomagnetic signal discernibility. TCN expands feature dimensions while preserving LSTM time-series characteristics.  The proposed stacking framework of multi-scale TCN and LSTM demonstrates effective indoor magnetic localization.
            DeepML, another smartphone-based indoor localization system, employs LSTM networks using magnetic and light sensors. Bi-modal images are generated via preprocessing and fed into the network for training, enabling new device localization. It's tested across varying environments, showcasing its consistency \cite{localization25}.
            
        In \cite{magneticnew2}, the authors introduce an indoor localization test-bed, achieving a high accuracy of 98$\%$ through XGBoost algorithms applied to magnetometer sensors and Wi-Fi access points, and examine classifier performance with varying test data sizes. Studied in \cite{magneticnew3,magneticnew4},  enhances smartphone indoor localization by combining Wi-Fi RSSI and magnetic field data through DNN. It encompasses offline learning to extract intrinsic features from multi-class fingerprints and an online serving phase. 
        Instead of Wi-Fi signals, \cite{magneticnew9} explored indoor localization using stable geomagnetic sensor signals. A DNN model and an  RNN track unique geomagnetic field signal sequences caused by object movement for positioning. Basic RNN and LSTM versions are trained on magnetic field maps of medium and large-scale indoor testbeds.
        Similarly, \cite{magneticnew5} utilizes smartphone sensors like magnetometer, accelerometer, and gyroscope for indoor localization. It employs fingerprinting based on magnetic flux intensity patterns, mitigating database updating and device heterogeneity issues. ANN aids user state identification (walking/stationary) with 95$\%$ accuracy. The study in \cite{magneticnew6} proposed a multi-sensor fusion approach to mitigate device dependency and enhance accuracy. A DCNN  recognizes indoor scenes to refine localization, while a magnetic field pattern database minimizes device reliance. Modified K nearest neighbor (mKNN), pedestrian dead reckoning, and an EKF further refine localization.  A CNN-based IPS utilizing magnetic patterns (MP) for localization is proposed in \cite{magneticnew7}. A database of MP is created, and CNN matches user-collected MP to estimate positions, employing a voting mechanism for accuracy. 
        An optimized geomagnetic positioning system using an enhanced genetic algorithm (EGA) and extreme learning machine (ELM) is proposed in \cite{magneticnew8}. EGA optimizes ELM's parameters, achieving meter-level accuracy with robustness and faster construction. All the papers mentioned in this section are analyzed in terms of complexity. scalability and cost in Table. \ref{magnetic}.
        However, magnetic sensor-based IPS encounters issues such as signal variability, device heterogeneity, limited range, dynamic environment effects, and the need for accurate real-time updates, impacting the system's accuracy and reliability.

        % \vspace{-0.2cm}
\subsection{Sensors-based}
			Sensors generate outputs proportional to the environmental conditions they are exposed to such as sound, temperature, pressure, light, etc., \cite{sensnw1}. Sensors can be widely divided into two categories, active sensors which can interact with the environment, and passive sensors which can only receive the data. Sensor-based indoor positioning systems consist of multiple sensors at predefined locations to detect and track the position of a person or a device \cite{sensnw2}. 
   The study in \cite{sensnw3} introduces a Cascaded DNN (CDNN) that utilizes smartphone sensor data to accurately localize objects in indoor environments, catering to the needs of individuals living alone or securing belongings.  The CDNN employs a tree structure of independent Deep Neural Networks (DNNs) for prediction. 
			   % \vspace{-0.2cm}

					\begin{table*}
      \renewcommand{\arraystretch}{1.1}	
		\caption{WSN-based indoor positioning systems $\&$ solutions. }
			\begin{tabular}{ | m{4.5cm} | m{4cm}| m{2cm} |  m{2cm} |  m{1.5cm} | m{1.5cm} | } 
			\rowcolor{lightgray}
			
			\hline
				% \vspace{.2cm}
			 Technology &  Positioning  algorithm  & Complexity & Scalability  & Cost & Paper \\ 
			\hline

			% \vspace{.2cm}
			 WSN & PNN &  Low & Medium & Low & 	\cite{wsn4}\\
			\hline
 % \vspace{.2cm}
		WSN & ANN   & Medium & Low & Low & 	\cite{wsn1} \\
			\hline
			
			% \vspace{.2cm}
			  WSN - Crossbow IRIS, PSO DV-Hop  & ANN &  Medium & Medium & Low & \cite{wsn2}, \cite{wsn3}\\
			\hline
			
				% \vspace{.2cm}
			 WSN & RBF &  Medium & Medium & Low & 	\cite{intro14}\\
			\hline

			% \vspace{.2cm}
			 WSN & DNN &  High & Low & Low& 	\cite{wsnnew3}\\
			\hline

			% \vspace{.2cm}
			 WSN & RBF-NN &  High & Medium & Low & 	\cite{wsn5}\\
			\hline

			% \vspace{.2cm}
			 WSN & Deep extreme learning machine &  High & Medium & Medium& 	\cite{wsnnew2}\\
			\hline

   % \vspace{.2cm}
			 Zigbee & kNN &  Low & Medium & Low & \cite{zigbeenew1}\\
			\hline 
     % \vspace{.2cm}
			  Zigbee & ANN &  Low & Medium & Low & \cite{zigbee1}\\
			\hline

    % \vspace{0.2cm}
   Zigbee & BP-ANN & Medium  & Low & Low & \cite{zigbeenew3} \\ \hline

	% \vspace{.2cm}
			 Zigbee & QPSO-GRNN &  Medium & Medium & Low & \cite{zigbee2}\\
			\hline 
       % \vspace{.2cm}
			 Zigbee & Weighted Nearest Algorithm &  Medium & High & Low & \cite{zigbeenew2}\\
			\hline

		\end{tabular}
  \label{wsn}
	\end{table*}

			\subsection{Wireless Sensor Networks (WSN)-based}

Wireless Sensor Networks (WSNs) interconnect sensors for tracking, surveillance, and intrusion detection, eliminating wired connections. They're vital in areas like forest fire-prone zones and high-risk areas where wired networks are impractical. Accurate location data for sensor nodes, along with transmitted or received data, ensures WSN efficiency. Despite traditional methods' shortcomings, new techniques using TDOA data with neural networks improve node location accuracy \cite{wsn1}. This employs two ANN models: Back Propagation Network (BPN) and Radial Basis Function (RBF), extracting distance information between anchor and sensor nodes. RBF outperforms BPN in accuracy, enhancing the system's precision in determining node locations.

			A fingerprint-based indoor localization approach using WSN is proposed in \cite{wsn2}. RSSI data collected from a real environment is used to train the feed-forward ANN to improve accuracy. The system performs better on statistical data than raw data. Similarly, an algorithm utilizing deep learning, extreme learning machines, and high-level features extracted by an autoencoder to enhance localization performance is proposed in \cite{wsnnew2}. Additionally, the method gradually increases the training data to update the fingerprint database and improve performance. 
			Since the wireless channels are vulnerable to various factors such as multipath effects and shadowing, modeling a proper propagation loss channel might be difficult and thus the RSSI-based positioning approaches can be erroneous. To avoid all these limitations, \cite{wsn3} proposed an indoor positioning technique based on NN and grid sensor training phase to localize the sensors accurately. The RSS data from the grid is used to train the NN. The estimation accuracy is directly proportional to the grid density and it is reliant on the variation in RSS data. An optimization-aided DNN for location prediction using distance-based features like AoA and RSSI is proposed in \cite{wsnnew3}. In this work, a hybrid model, the Lion Assisted Firefly Algorithm (LAFA), enhances localization accuracy. The study presents a parametric analysis of the LAFA algorithm, assessing performance and variations.
			S. Rajaee $ et ~ al.$ \cite{wsn4} addresses a positioning approach in ad-hoc wireless networks, where the nodes and anchors are uniformly distributed in a squared area. The approach uses a probabilistic neural network (PNN) for estimating the location of unknown nodes. The proposed system uses  Independent Component Analysis (ICA) to reduce the dimension of the data and to remove the unwanted data which helps to simplify the data processing. By the use of ICA, the energy consumption is reduced by 43$\%$. The minimum positioning error for the system is 16$\%$.
		Guangzhu $ et ~ al.$ proposed a WSN-based localization technique for tracking moving targets in real time in \cite{wsn5}. This system works on predicting the location of the moving target in chain-type WSNs using an RBF neural network. The target tracking prediction problems perform well in real-time. The major shortcoming of this work is that the prediction algorithm will perform well in single-target tracking problems. The performance of the system in multi-target tracking prediction is limited. A similar system is proposed in \cite{intro14}. This study explores NNs for solving localization issues in noisy distance measurements in WSNs. Comparing MLP, RBF and RNN, alongside KF variants, it finds that RBF offers the best accuracy but the highest resource requirements.

 The study in \cite{wsnnew1} explores device-free wireless localization and activity recognition, estimating a person's location and activity without equipping them with a device. It uses radio image processing to characterize human behavior influence on Wi-Fi signals. Unlike traditional methods, it considers correlated CSI measurements from multiple channels as a radio image. Deep learning is employed to extract optimized deep features from image features, achieving excellent performance in location and activity estimation in cluttered laboratory experiments.

 % {\color{red}\cite{wsnnew1,wsnnew2,RF3}}

			\subsubsection{Zigbee-based}

   Zigbee, a decentralized mobile ad-hoc network, finds application in low-power, low-data-rate scenarios such as home automation, data collection, and short-range data transfer. Operating at a data transfer rate of approximately 250 kbits/sec, Zigbee's strength lies in its energy efficiency and long battery life, albeit with a range limited to 10-100 meters. Its secure networking is facilitated by 128-bit symmetric encryption keys. Various studies explore Zigbee's potential in indoor localization, primarily utilizing the RSS fingerprinting method with Artificial Neural Networks (ANNs). Optimization algorithms like Quantum Particle Swarm Optimization (QPSO) are integrated, enhancing localization precision. Zigbee-assisted indoor localization (ZIL) enhances Wi-Fi-based positioning by capturing mixed Wi-Fi signals, employing novel fingerprint-matching algorithms and K Nearest Neighbors (KNN) techniques for accuracy improvement. Additionally, location calculations in certain scenarios employ algorithms such as nearest, weighted nearest, and Bayesian algorithms, yielding enhanced accuracy over longer distances. This approach is particularly beneficial for larger indoor spaces or underground environments, augmenting location-based services. In a distinct application, a Wireless Sensor Network (WSN) coupled with Back Propagation Artificial Neural Network (BP-ANN) is utilized for indoor localization of Alzheimer's patients, demonstrating the versatility of Zigbee technology in healthcare. A comparative analysis of these studies is provided in Table. \ref{wsn}.

 While WSN and Zigbee-based IPSs are widely adopted, they encounter notable challenges. Firstly, signal propagation is vulnerable to obstacles and environmental conditions, causing inaccuracies in distance measurements and positioning outcomes. Secondly, the constrained range and limited data transfer speed of Zigbee can undermine coverage and data transmission reliability. Thirdly, network congestion can emerge from the growing device count, impacting both communication quality and positioning precision.  These combined challenges collectively influence the resilience and effectiveness of IPSs based on WSN and Zigbee technologies.
	% \vspace{-0.2cm}	

   % \vspace{-0.2cm}
			
		\begin{table*}
   \renewcommand{\arraystretch}{1.1}	
		\caption{RFID-based indoor positioning systems $\&$ solutions. }
			\begin{tabular}{ | m{3.5cm} | m{3cm}| m{2cm} |  m{2cm} |  m{2cm} | m{3cm} | } 
			\rowcolor{lightgray}
			
			\hline

   	% \vspace{.2cm}
			 Technology &  Positioning  algorithm  & Complexity & Scalability  & Cost & Paper \\ 
			\hline

			% \vspace{.2cm}
			 RFID & ANN  & Low & Low & Low & \cite{rfid11}\\
			\hline

		 %\vspace{.2cm}
			 RFID& ANN  & Low & Medium & Low & \cite{rfid8}\\
			\hline
			
			 %\vspace{.2cm}
			  RFID & RBF  & Low & Medium & Low & \cite{rfid9}\\
			\hline
				 %\vspace{.2cm}
			  RFID & CMTL  & Low & Medium & Medium & \cite{rfid16}\\
			\hline
			 %\vspace{.2cm}
			 RFID & AIS-RBF-NN  & Low & High & Low & \cite{rfid2}\\
			\hline
			
			 %\vspace{.2cm}
		RFID & ANN & Low & High & Low &  \cite{rfid3}, \cite{rfid5}\\
			\hline

   		 %\vspace{.2cm}
		  RFID& ANN  &  Low & High & Low & \cite{rfid14}\\
			\hline

			 %\vspace{.2cm}
			  Active RFID - RSS &  kNN  & Medium &  Low & Low & \cite{knn1}\\ 
			\hline

    %\vspace{.2cm}
			 RFID & ANN   & Medium & Medium & Low & \cite{rfid10}\\
			\hline
			 %\vspace{.2cm}
			 RFID & PSO-ANN & Medium & High & Low & \cite{rfid6}\\
			\hline
			
			 %\vspace{.2cm}
			 RFID & ANN &  Medium & High & Low & \cite{rfid1},\cite{rfid12}\\
			\hline

			 %\vspace{.2cm}
			  RFID & RBF  & Medium & High & Low & \cite{rfid13}\\
			\hline

			 %\vspace{.2cm}
			 RFID- RSSI & BP-ANN & High & Medium & Low & \cite{rfid7}\\
			\hline

		\end{tabular}
  \label{rfid}
	\end{table*}

			% \vspace{-0.2cm}
			\subsection{Radio Frequency Identification (RFID)-based}
			
An RF-compatible circuit facilitates electromagnetic transmission for data storage and retrieval. Essential components of a basic RFID system include RFID readers, RFID tags, and elements for communication. The system transmits and receives data within a predetermined radio frequency and protocol, operating in either passive or active mode \cite{RFID1}.

\par Passive RFID systems serve as alternatives to traditional bar code technology, being smaller, simpler, and more cost-effective compared to active systems. These tags operate without a battery, reflecting the RF signal from the reader to the receiver, with information added by modulating the reflected signal. However, their operational range is limited to 1-2 meters, making them unsuitable for larger areas. Commonly used frequency bands include LF, HF, UHF, and microwave frequencies \cite{tdoa1}. Active RFID systems utilize similar frequency ranges, featuring small transceivers that communicate their ID upon interrogation. These systems typically offer a larger range and find applications like SpotON, a 3D location sensing system utilizing radio signal strength analysis for object location detection \cite{RFID3}.

In \cite{rfid6}, a Particle Swarm Optimization Artificial Neural Network (PSO-ANN) algorithm for RFID indoor positioning is proposed, using PSO to optimize ANN weights and thresholds. This cost-efficient RFID IPS is compared with BPANN, ANN, and LANDMARC models. The approach establishes the relationship between RSSI and tag position, enhancing accuracy. A Gaussian filter is employed for data processing, mitigating environmental effects. Achieving an average positioning error of 0.6482 m, the method leverages non-contact two-way communication for data transfer. In a similar vein, \cite{rfid7} presents a Genetic Algorithm-Backpropagation Neural Network approach, combining GA's optimum searching with BPNN's optimization for RSSI-based indoor positioning. Also, \cite{rfid8} introduces a high-accuracy indoor location system using active RFID and neural network classification. In \cite{rfid9}, a Radial Basis Function Neural Network (RBFNN) with Virtual Reference Tags is employed for RFID-based target localization, achieving a consistent accuracy with a positioning error of 0.472 m, while selecting optimal network architectures.

In \cite{rfid10}, an integrated wireless platform proposes an adaptive RFID-derived RSSI-based indoor location sensing technique. Utilizing a fuzzy neural network architecture, the method adapts to environmental parameters. Active readers and tags are used to enable long-distance transmission, achieving less than 1 m positioning error with fewer tags and readers. The approach in \cite{rfid11} introduces an intelligent tag strength prediction algorithm using backpropagation learning in neural networks for RFID tag position detection. It predicts tag signal strength under varying conditions and achieves over 90$\%$ accuracy in estimating target positions. While \cite{rfid12} presents an RFID hybrid positioning method employing a neural network phased array antenna for indoor RF localization. Combining AOA and RSSI, it scans the search plane using phased array antenna radiation beams, achieving a mean positioning error of 0.32m for 10 locations.

In \cite{rfid13}, an algorithm is presented for estimating target location in dynamic indoor environments like warehouses with changing layouts. By utilizing RSSI and passive UHF tags as references, a trained RBFNN reduces the Localized Generalization Error (L-GEM) for 2D warehouse positioning. Similarly, \cite{rfid14} employs multiple neural networks and a genetic algorithm to estimate indoor positions from RSSI data collected from reference tags, achieving an average error of 2.4 m and a maximum error of 5.21 m. Additionally, \cite{rfid16} explores active RFID for positioning moving targets using Cluster-based Movable Tag Localization (CMTL) with kNN and ANN, achieving an average positioning error of 0.77 m. These studies are compared in Table \ref{rfid}. Despite numerous approaches for IPSs utilizing RFID technology, challenges persist, including fluctuations in signal strength, positioning inaccuracies, environmental factors, and tag-related configurations, underscoring the need to address these issues for improved reliability and accuracy.

% \vspace{-0.2cm}

	\begin{table*}[!h]
  \renewcommand{\arraystretch}{1.1}	
		\caption{Bluetooth-based indoor positioning systems $\&$ solutions.}
		\begin{tabular}{ | m{3.5cm} | m{3.5cm}| m{2cm} |  m{2cm} |  m{2cm} | m{2.5cm} | } 
			\rowcolor{lightgray}
			\hline
   	 %\vspace{.2cm}
						 Technology &  Positioning  algorithm  & Complexity & Scalability  & Cost & Paper \\ 			\hline
        % %\vspace{.2cm}

		 Bluetooth &  RF & Low & Low& Low  &	\cite{bluetooth14}\\
			\hline
			    %\vspace{.2cm}
			Bluetooth & RF  &  Low & Medium & Low & \cite{zigbee3}\\
			\hline

	 %\vspace{.2cm}
			  Bluetooth -RSSI & Denoising Auto Encoder&  Low & High & Low & \cite{localization33} \\
			\hline

  %\vspace{.2cm}
			 Bluetooth & KNN, SVM, NN  & Medium & Low & Low &  \cite{bluetooth8}\\
			\hline
			
				 %\vspace{.2cm}
			 Bluetooth & RNN &  Medium & Low & Low &  \cite{bluetooth6} \\
			\hline

       %\vspace{.2cm}
		 Bluetooth  & CNN, PCA, t-SNE  & Medium &  Low & Low &	\cite{bluetooth13}\\
			\hline

    %\vspace{.2cm}
		 Bluetooth &  NN &  Medium & Low & Low&	\cite{bluetooth15}\\
			\hline

   		 %\vspace{.2cm}
			 Bluetooth & ANN &  Medium & Medium &Low & \cite{bluetooth4,bluetooth5}\\
			\hline

 %\vspace{.2cm}
		 Bluetooth-RSS &  XGBoost  & Medium & Medium& Low &	\cite{bluetooth11}\\
			\hline

   	   %\vspace{.2cm}
		 Bluetooth  &  Weighted Nearest Neighbors  & Medium & High & Low &	\cite{bluetooth10}\\
			\hline
   %\vspace{.2cm}
		 Bluetooth-RSS, AoA &  CNN-PCA & High & Low & Low &	\cite{bluetooth12}\\
			\hline
   
  		 %\vspace{.2cm}
		 Bluetooth-RSS &  CNN  & High & High & Low &	\cite{bluetooth9}\\
			\hline

		\end{tabular}
  
  \label{ble}
	\end{table*}
	% \vspace{-0.2cm}
\subsection{Bluetooth-based}
	
Bluetooth, a wireless technology used for short-range communication among various devices, like computers and smartphones, employs affordable transceiver microchips and radio systems, ensuring minimal power consumption. Its operational range, typically around 10-15 meters, varies based on factors such as materials and has a widespread standard operating frequency of 2.4 GHz with a lower bit rate. Bluetooth-based indoor localization, favored for its presence in mobile devices, cost-effectiveness, and energy efficiency, utilizes unique IDs for precise tag location \cite{bluetooth3}, proving valuable for tracking and room-specific tasks.
   
			\par
The Denoising Autoencoder-based Bluetooth Low Energy (BLE) indoor localization (DABIL) \cite{localization33} employs a 3D indoor positioning system utilizing Bluetooth data. Utilizing a denoising autoencoder, it extracts relevant fingerprints from RSSI data to create a 3D reference point database. This method demonstrates improved vertical and horizontal accuracies with a 1.27m positioning error. Neural networks are commonly utilized for Bluetooth-based localization. For instance, in one approach \cite{bluetooth4}, a cost-effective system utilizes neural networks for user orientation, achieving a 0.5m accuracy. Another study \cite{bluetooth5} employs Bluetooth in phones for neighboring positions, using a 6-neuron hidden layer with a 17$\%$ error rate. Recurrent networks in \cite{bluetooth6} focus on real-time Bluetooth localization with a 10m error. In \cite{bluetooth8}, kNN surpasses neural networks and SVM for indoor Bluetooth positioning with less than 1m error. Bluetooth RSS in \cite{bluetooth9}, employing CNNs, attains 93.33$\%$ accuracy and less than 1.4m error, suitable for multi-floor settings, effectively covering large areas.

            The study in \cite{bluetooth10} proposes a BLE-based indoor localization system utilizing Android devices to create continuous radio maps with BLE "iBeacons" and Wi-Fi access points. Stationary object localization is assessed using Wi-Fi, BLE, and their combination, with optimal parameter selection through Weighted Nearest Neighbors (WNN). Another study in \cite{bluetooth7} proposes a remote indoor positioning system using kNN analysis and a portable BLE tag, applicable to tracking individuals in various scenarios. 
            The demand for precise indoor localization arises from superstores, smart homes, and disaster management needs. Addressing the complexity of indoor settings, \cite{bluetooth11} describes an explainable indoor localization (EIL) technique employing BLE's RSSI with a gradient boosting machine. It achieves 98.04$\%$ accuracy within 1.5m in a superstore environment. 

            This study in \cite{bluetooth12} explores IoT-driven location-based services, particularly in indoor Bluetooth localization using Bluetooth 5.1's AOA function. They proposed a DL-based algorithm that fuses RSSI and AOA features through PCA and KF. CNN extracts deep-level features from RSSI and AOA, followed by concatenation and Softmax layer classification. The authors of  \cite{bluetooth13} also proposed an indoor localization system using Bluetooth fingerprinting and CNN. A unique approach transforms wireless signal data into images using a blurring technique to simulate signal diffusion. Additionally, two-dimensional reduction algorithms, PCA and t-SNE, are compared. An evolutionary algorithm configures the solution with varied transmission power levels. Results demonstrate a promising accuracy of nearly 94$\%$, highlighting the potential of this technique for enhanced indoor localization systems.

            An improved RSSI-based fingerprinting method, employing data augmentation and ML algorithms for XY-position identification of user nodes relative to anchor nodes, is proposed in \cite{bluetooth14}. RF achieved a 96$\%$ test accuracy, outperforming other techniques, ensuring precision, and compatibility with ML. In \cite{bluetooth15}, a smartphone-based indoor location method utilizing BLE Beacons' RSSI values is suggested. ML is used to create a distance estimator from RSSI readings. TensorFlow is employed to calculate intersection points of peripheral lines, facilitating position estimation based on the geometric median. Additionally, \cite{zigbee3} enhances accuracy by utilizing multiple anchors and radio channels. RF proves most effective, achieving over 99$\%$ classification accuracy. These methods are compared in Table.\ref{ble}. Despite Bluetooth technology's energy efficiency, device compatibility, and cost-effectiveness in IPS, challenges like limited range, signal interference, accuracy variation, and privacy concerns persist. Overcoming these challenges necessitates careful implementation and innovative solutions for optimal performance in indoor positioning systems.

			% \vspace{-0.2cm}
			\subsection{Wi-Fi-based}

	\begin{table*}
  \renewcommand{\arraystretch}{1.1}	
		\caption{Wi-Fi-based indoor positioning systems $\&$ solutions. }
			\begin{tabular}{ | m{3.5cm} | m{3.6cm}| m{2cm} |  m{2cm} |  m{2cm} | m{2.4cm} | } 
			\rowcolor{lightgray}
			
			\hline
				 %\vspace{.2cm}
			 Technology &  Positioning  algorithm  & Complexity & Scalability  & Cost & Paper \\ 
			\hline

					 %\vspace{.2cm}
		  RF- RSSI & ANN & Low & Low & Medium & \cite{wsn2} \\
			\hline
          % %\vspace{.2cm}			
		Wi-Fi-RSSI &Bayesian Network& Low&Low&Medium& \cite{wlan7}\\
            \hline
			 %\vspace{.2cm}
	       
			 Wi-Fi -RSS & DNN,LDA & Low & Medium & Low & \cite{localization8}\\
			\hline

			 %\vspace{.2cm}
			 Wi-Fi-5g & DNN\ & Low & Medium & Low &  \cite{localization19} \\
			\hline

			 %\vspace{.2cm}
			 Wi-Fi & DNN  & Low & Medium & Low & \cite{localization20, wifinew16}\\
			\hline

% 				 %\vspace{.2cm}
% 			Wi-Fi fingerprint & CNN & High & High & Medium & \cite{new2}\\
% 				\hline
 %\vspace{.2cm}
			 Wi-Fi-RSSI & Bayesian location estimator &  Low & Medium & Low &  \cite{wlan6}\\
			\hline

		 %\vspace{.2cm}
   Wi-Fi - CSI & PCNN & Low&Medium&Low&\cite{wifinew11} \\
    \hline

     %\vspace{.2cm}			
		Wi-Fi &ANN&Low&Medium&Low& \cite{wlan17}	\\
            \hline
             	 %\vspace{.2cm}
 %\vspace{.2cm}
		  Wi-Fi- Signal Strenth  & kNN, Viterby-like algorithm &  Low & High & Low & \cite{wlan3,Rad1} \\
			\hline
   
		 	  %\vspace{.2cm}
    Wi-Fi RSSI&PCA&Low&Medium&Medium&\cite{wifinew12} \\
    \hline

   	 %\vspace{.2cm}
			 Wi-Fi-RSSI & Bayesian network &  Low & Medium & Medium & \cite{wlan11}  \\
			\hline
       % %\vspace{.2cm}
			 Wi-Fi - CSI & RF  & Low & Medium & High & \cite{wifinew7} \\
			\hline

			 %\vspace{.2cm}
			  Wi-Fi-RSSI  & SVM &  Low & Medium & High &  \cite{svr2}\\
			\hline

            	 %\vspace{.2cm}
			 Wi-Fi, Magnetic Field & DNN  & Medium &  Low& Low  & \cite{localization14}  \\
			
			\hline

     %\vspace{.2cm}			
		Wi-Fi&DNN&Medium&Low&Medium& \cite{wlannew1}	\\
            \hline
               % %\vspace{.2cm}
			 Wi-Fi - CSI & 1D-CNN  & Medium & Low & Medium & \cite{wifinew6} \\
			\hline
			 %\vspace{.2cm}	
		Wi-Fi - RSSI & CNN   & Medium & Medium & Medium  & \cite{localization11}, \cite{new2}\\
			\hline
  %\vspace{.2cm}
			  Wi-Fi- CSI & CNN &  Medium & High & Low & \cite{localization30, new6, wifinew9}\\
			\hline

	 %\vspace{.2cm}
		  Wi-Fi - RSSI  & kNN  & Medium & High & Low &  \cite{wlan4, wlan5}\\ 
			\hline
	 %\vspace{.2cm}
			Wi-Fi - RSSI & MMLP & Medium & High &Low &\cite{intro18} \\     \hline
			%  %\vspace{.2cm}
 %\vspace{.2cm}
			 Wi-Fi - RSS &  MLP, SVM  & Medium& High & Low &\cite{intro17, svr1} \\ 
			\hline

               % %\vspace{.2cm}

               Wi-Fi - RSSI & kNN & Medium&High &Medium&\cite{wlannew2} \\
                \hline
			
			 %\vspace{.2cm}
			 Wi-Fi-RSSI & Joint Clustering  & Medium & High & Medium &  \cite{wlan4,wlan5} \\
			\hline

				 %\vspace{.2cm}
			Wi-Fi fingerprint & D-CNN & Medium & High & Medium & \cite{new2}\\
				\hline

							 %\vspace{.2cm}
			 Wi-Fi - RSSI & DNN & Medium & High&  Medium &  \cite{localization10,localization3}  \\
			\hline

     %\vspace{.2cm}
    Wi-Fi RSS& D-CNN,SVM&Medium&High&Medium&\cite{new7} \\
    \hline

		    %\vspace{.2cm}
			 Wi-Fi - CSI & CNN  & Medium & High & Medium & \cite{wifinew5} \\
			\hline

	 %\vspace{.2cm}
			%\bibitem{localization28}
			Wi-Fi MIMO channel & CNN & Medium & High & High & \cite{localization28} \\

			\hline

    		 %\vspace{.2cm}
     Wi-Fi - CSI&1-DCNN-LSTM&High&Low&Low&\cite{trajec1}\\
    \hline	

    %\vspace{.2cm}
			 Wi-Fi - CSI & DL  & High & Low & Medium & \cite{wifinew3} \\
			\hline

    %\vspace{.2cm}
			 Wi-Fi - CSI - MIMO & DNN  & High & Low & High & \cite{wifinew4} \\
			\hline

		 %\vspace{.2cm}
		 Wi-Fi,LTE,Magnetometer data & pattern matching algorithm &  High & Medium & Low & \cite{wifinew2}\\
			\hline
   
			 %\vspace{.2cm}
			  Wi-Fi-RSSI & DNN, Auto-encoder &  High & Medium & Medium & \cite{localization22}\\
			  \hline
			  
			   %\vspace{.2cm}
			Wi-Fi-RSS &	DNN-CNN-DS & High & Medium & Medium & \cite{new3} \\
				\hline
   
     %\vspace{.2cm}
    Wi-Fi -RSSI & DNN, KF&High&Medium & Medium&\cite{localization21} \\
    \hline

         % %\vspace{.2cm}
			 Wi-Fi - CSI &  DNN   & High &Medium & Medium & \cite{localization6, wifinew1, wifinew10}\\ 
			\hline

			 %\vspace{.2cm}
			Wi-Fi-RSS &	LDA, SVM, KNN, RF & High & Medium & Medium & \cite{new5} \\
				\hline
    	 %\vspace{.2cm}
			 Wi-Fi - CSI & DCNN  & High & High & Low & \cite{localization16,wifinew8} \\
			\hline

					    %\vspace{.2cm}

    Wi-Fi-RSSI & PaCNN-LSTM & High &High&Medium&\cite{wifinew15}\\ 
    \hline	
    % \vspace{.2cm}

		\end{tabular}
		\label{wifin}
	\end{table*}
	
Wi-Fi, a widely adopted wireless technology, links devices to the internet via routers, offering a range of 20 to 150 meters, extendable through overlapping Access Points (APs), making it ideal for localization systems due to its global presence \cite{new7}. Its speed and range adhere to IEEE protocol standards, with localization relying on ranging-based and fingerprinting-based methods. WLAN, a local wireless network, operates in limited areas like schools and campuses, using technologies such as Frequency Hopping Spread Spectrum (FHSS) and Direct Sequence Spread Spectrum (DSSS). APs, acting as routers, connect to the internet, typically utilizing the 2.4 GHz frequency with a range of 50-100 meters. Wi-Fi and WLAN-based IPS leverage these technologies for precise and dynamic localization within enclosed spaces, revolutionizing indoor navigation.  By capitalizing on the proliferation of access points, routers, and beacons, these systems enable the tracking and positioning of individuals and objects in real-time, revolutionizing how we navigate and interact with indoor spaces. While Wi-Fi and WLAN are frequently used interchangeably, in this context, we are addressing them under the  Wi-Fi-based category. 
       
        \subsubsection{Wi-Fi Fingerprinting-Based }
In fingerprinting-based methods, creating a fingerprint map and matching it to online fingerprint data requires significant computing resources. However, indoor scenarios' complexity often leads to poor data quality during offline data collection for fingerprint databases. Therefore, employing ML in fingerprint-based indoor positioning is preferable, reducing computing resource usage without sacrificing accuracy. Additionally, this approach facilitates floor distinction.      
        
        \paragraph{RSSI-Based}
        
RSSI-based fingerprinting in Indoor Positioning Systems (IPS) serves to estimate the location of mobile devices within indoor environments by measuring the power level of Wi-Fi signals received from nearby Access Points (APs) or routers. While offering simplicity, low-cost implementation, and compatibility with most Wi-Fi-enabled devices, RSSI-based methods are susceptible to environmental changes like signal interference, multipath effects, and signal strength variations due to obstacles. Nonetheless, RSSI-based fingerprinting remains widely used, especially in environments where additional infrastructure installation isn't feasible, such as public spaces and shopping malls \cite{localization3}. Recent years have seen an exploration of Deep Neural Networks (DNNs) in RSSI-based IPSs, with systems utilizing DNNs to classify and regress Wi-Fi user positions in indoor environments \cite{localization8, localization14, localization10, wifinew16}. For instance, A. Adege et al. introduced a DNN-based indoor localization technique employing Wi-Fi \cite{localization8}, achieving high accuracy by preprocessing RSS data with Linear Discriminant Analysis (LDA) and selecting the strongest RSS values. Other techniques like Deep Positioning \cite{localization14} combine RSSI and magnetic field data for improved accuracy, while scalable approaches like the one proposed by K. S. Kim et al. \cite{localization10} utilize hierarchical DNN architectures for multi-floor classification with consistent performance. Additionally, in \cite{localization11}, a Wi-Fi-based approach for localization of Mobile Nodes (MNs) is introduced, employing RSSI to create location fingerprints and employing a Convolutional Neural Network (CNN) on time series data to enhance accuracy, achieving a mean error of 2.77m in predicting coordinates.

In \cite{wlan6}, Roos $ et ~ al.$ proposed an IPS. This grid-based Bayesian estimator achieved over 50$\%$ accuracy within 1.5 meters in a small office. Similarly, \cite{wlan7} employed a Bayesian probabilistic method for device localization. Battiti $ et ~ al.$ in \cite{intro17} developed a system using an MLP-based classifier and OSS training, yielding less than 3 meters error with only 5 signal strength samples. Its advantage lies in lower sensitivity to overfitting. \cite{wlan9} compares neural network and nearest neighbor classifiers, achieving 72$\%$ accuracy within 1 meter. An MLP in \cite{intro17} maps RSS data to user location with an average accuracy of 2.3 meters, leveraging Wi-Fi infrastructure while addressing privacy. \cite{intro18} introduces the Modular Multi-Layer Perceptron (MMLP) for enhanced accuracy in location estimation, managing uncertainty factors.

In \cite{wlan4,wlan5}, the Horus system employs a joint clustering technique for probabilistic location estimation. Each person's location coordinate is treated as distinct classes, and $ L_{i}$ is selected based on the highest likelihood to minimize distance error. The experiments achieved over 90$\% $ accuracy within 2.1 meters, with improved accuracy observed as the number of samples per location increased. \cite{svr1} presents an indoor positioning method using SVM and statistical learning theory, showcasing SVM's low error rate as a classifier and its regression version for mobile user positioning. Another approach in \cite{svr2} determines device location using RSSI from APs, employing an SVM-based fingerprinting algorithm for real-time analysis. \cite{intro19} introduces the Discriminant-Adaptive Neural Network (DANN) for Wi-Fi client positioning, utilizing RSS from APs to construct an accurate RSS-position relationship. Additionally, \cite{wlan17} proposes an ANN-based approach for real-time target location detection and room type identification with median positioning errors of 5.46m and 3.75m respectively. Finally, \cite{wlannew4} presents a hybrid WiFi and WSN-based approach using ANN, achieving an average distance error reduction to 1.05 meters, surpassing GA optimization.

In their work \cite{wlan11}, Ladd $ et ~ al.$ presented a grid-based Bayesian algorithm employing the 802.11 standard for robot localization. Using RSS data from 9 APs, the host employs a probabilistic model to compute the position likelihood from a pool of locations, refining the results by considering the mobile host's limited maximum speed. Achieving over 83$\% $ accuracy within 1.5 meters, this method proves effective for robot localization. Similarly, \cite{wlan12} adopts a practical Bayesian approach with the 802.11 architecture for topological localization within office buildings, reducing training time without sacrificing accuracy.
J. Zou et al. proposed a Wi-Fi localization system \cite{new3} utilizing RSS with a deep regression model named DNN-CNN-DS, comprising DNN, CNN, and Dempster-Shafer. An Auto-Encoder initializes the DNN weights, optimized by minimizing the mean square error between model output and real location. DFLAR, a device-free wireless localization and activity recognition technique \cite{localization22}, employs deep learning to recognize activities and gestures based on the target's influence on nearby wireless links. Similarly, the approach in \cite{wifinew15} enhances indoor Wi-Fi localization accuracy by approximately 22$\% $ through improved contrastive learning and a parallel fusion network, PaCNN-LSTM. Another system in \cite{wifinew12} enhances WiFi indoor localization efficiency and performance by utilizing PCA.

Another popular strategy in fingerprinting for indoor positioning involves utilizing fingerprint images derived from RSSI data. Numerous approaches based on fingerprint image processing and deep learning are presented in \cite{new7,new2,new5}. In \cite{new7}, a dilated CNN is trained on RSS images, and prediction errors are utilized to train an SVR model, verified using the UJIIndoorLoc dataset \cite{dataset1}. Alternatively, in \cite{new2}, Wi-Fi and magnetic signals are transformed into fingerprint images, with a CNN learning mappings to actual positions, showcasing robust learning and accuracy. Similarly, \cite{new5} introduces MFMCF, leveraging multi-pattern fingerprints and various classifiers (KNN, SVM, RF) to enhance localization accuracy by constructing a composite fingerprint set (CFS) with LDA from SSD, HLF, and RSS. Additionally, \cite{wifinew2} presents Wi-LO, an indoor localization system that boosts Wi-Fi-based accuracy by integrating LTE and magnetometer data, overcoming mismatches by combining different data types at each location. Furthermore, another study \cite{wlannew2} proposes an ML framework employing Bag-of-Features and kNN classification, surpassing existing models in both simulations and real-time experiments.

        \paragraph{CSI-Based}
        CSI-based fingerprinting in IPS utilizes detailed information from Wi-Fi signals' CSI to create distinct fingerprints for various indoor locations. This approach offers high accuracy and resilience in complex indoor environments but requires precise calibration and dense reference point deployment \cite{localization6}. Recent studies have extensively employed DL for CSI-based fingerprinting in IPS \cite{wifinew14,wifinew3,wifinew4,wifinew5,wifinew6}, providing automatic feature learning, noise resilience, adaptability to new environments, and real-time processing \cite{wifinew7,wifinew8,wifinew9,wifinew10,wifinew11}. Combining Wi-Fi signals with other methods can enhance accuracy, as seen in \cite{localization21,localization20}, where autoencoders reduced data dimensions, aiding in position estimation. Additionally, systems like BiLoc \cite{localization19} and \cite{localization16} leverage DL to estimate location using 5-GHz Wi-Fi CSI, achieving promising accuracy. Meanwhile, \cite{wlannew1} introduces data rate (DR) fingerprinting for passive localization, addressing challenges like low resolution and fluctuations by employing various strategies such as transmission power levels and dynamic nearest neighbors matching.
        
In \cite{localization28}, researchers explore utilizing Massive MIMO channels with CNN for localization, leveraging the sparse structure of these channels to achieve fractional wavelength positional accuracy. Another method, ConFi, employs CNN for Wi-Fi localization \cite{localization30}. It treats time-frequency metrics from CSI data as images and utilizes a 5-layer CNN for classification, demonstrating robust performance with a 2.7 m localization error. Additionally, a CSI image-based indoor localization method is introduced in \cite{new6}, forming an RGB image with phase differences and amplitudes from different antennas, and employing a CNN for classification. Another CSI-based indoor fingerprinting system, DeepFi, trains all DNN weights as fingerprints using deep learning in the offline phase, employing a greedy algorithm for complexity reduction. In the online phase, a probabilistic data fusion method based on the RBF is utilized for location estimation.

        \subsubsection{Wi-Fi Ranging-Based }
        A Wi-Fi IPS that utilizes trajectory CSI observed from predetermined routes instead of stationary locations, addressing the limitations caused by multipath fading is proposed in \cite{trajec1}. The proposed IPS employs a one-dimensional convolutional neural network-long short-term memory (1DCNN-LSTM) architecture to leverage the spatial and temporal information of trajectory CSI. Additionally, a generative adversarial network (GAN) helps enlarge the training dataset, reducing the cost of trajectory CSI collection. 
        All these studies are analysed in Table. \ref{wifin}.

	\begin{table*}
 \renewcommand{\arraystretch}{1.1}	
		\caption{Cellular and UWB-based indoor positioning systems $\&$ solutions. }
			\begin{tabular}{ | m{3.5cm} | m{3cm}| m{2cm} |  m{2cm} |  m{1.5cm} | m{3.5cm} | } 
			\rowcolor{lightgray}
			
			\hline
				 %\vspace{.2cm}
			 Technology &  Positioning  algorithm  & Complexity & Scalability  & Cost & Paper \\ 
			\hline
			 %\vspace{.2cm}

			  Cellular  & SVM, NN    & Medium  &  Medium & High & \cite{new8} \\ 
			\hline
     %\vspace{.2cm}
			  Cellular  & kNN    & Medium  &  High & Medium & \cite{knn2} \\ 
			\hline

			 %\vspace{.2cm}
		 Cellular & DNN & Medium & High & High & \cite{cell5},\cite{cell4}, \cite{cell6},\cite{cell7},\cite{cell8}\\
			\hline

    %\vspace{.2cm}
   Cellular -5G & SVM,KF & High & Medium & High & \cite{5gnew1}\\ \hline
     %\vspace{.2cm}
   Cellular -5G & DNN & High & High & High & \cite{5gnew2}\\ \hline

	 %\vspace{.2cm}
			 UWB & MLP, RPF &  Low & Low & High &  \cite{uwb5}\\
			\hline

			 %\vspace{.2cm}
			 UWB & ANN &  Low & Medium  & High & \cite{uwb2}, \cite{uwb3}\\
			\hline
				 %\vspace{.2cm}
			 UWB & CNN & Medium & Low & High & \cite{uwb6,uwb8}\\
			\hline

       % %\vspace{.2cm}
			 UWB & PCA, MC-SVM  & Medium& Medium & High  &  \cite{uwbnew1}\\
	\hline

			 %\vspace{.2cm}
			 UWB- RSSI & ANN  & Medium & High & High &  \cite{uwb4}\\
			\hline

        % %\vspace{.2cm}
			 UWB & LSTM& Medium & High & High &  \cite{rtt2}\\
    	\hline

     		 %\vspace{.2cm}
	
			 UWB & PNN & High & Low & High &  \cite{uwbnew3}\\
    	\hline
				
      % %\vspace{.2cm}
			 UWB & DNN - LSTM & High  & Medium & High &  \cite{rtt3}\\
    	\hline
			
	 %\vspace{.2cm}
			 UWB & CNN & High & Medium & High & \cite{uwb7}\\
			\hline

		\end{tabular}
		\label{wlancell}
	\end{table*}
	
 Wi-Fi signals usually encounter several challenges including signal propagation obstruction, multipath effects, and non-line-of-sight scenarios, leading to inaccuracies. Achieving high accuracy is difficult due to dynamic environments, infrastructure dependency, and privacy concerns. Calibration and maintenance are crucial, and interference in crowded channels can impact accuracy. Implementation complexity, cost, and power consumption pose further issues, requiring advanced algorithms, hardware improvements, and robust signal processing techniques for reliable Wi-Fi-based IPSs.
        % \vspace{-0.2cm}

\subsection{Cellular}
Localization methods utilizing cellular data encounter limited popularity due to intense multipath noise and setup heterogeneity, leading to signal strength variation and low accuracy within a range of 50-200 meters \cite{Cell1, Cell2}. The concept proposed in \cite{Cell2} involves employing wide signal strength fingerprints for indoor localization, incorporating the six strongest GSM cells and up to 29 additional channel readings. While this enhances accuracy through increased dimensionality, such channels may hinder efficient communication. Moreover, this technique is adaptable to IS-95 CDMA and 3G technologies. Another study \cite{cell5} introduces a device-agnostic fingerprinting-based localization method using deep learning, ensuring consistent performance with minimal power usage and enhancing robustness and system transparency.

SoloCell, described in \cite{cell6}, employs deep learning to optimize indoor positioning using signal strength history from cell towers. With 7 data collection modules, its performance was evaluated across diverse Android devices. MonoDCell \cite{cell7}, another cellular-based method, enhances data collection efficiency and robustness through deep LSTM networks and multiple modules, maintaining accuracy within 0.95 to 1.4 meters across different Android implementations. Additionally, OmniCells \cite{cell8} utilizes auto-encoders for device-invariant RSS data learning, ensuring consistent median localization accuracy even for unknown devices. CellinDeep \cite{cell3} utilizes DL for nonlinear cellular data-location relations, ensuring medium accuracy and 93.45$\%$ power efficiency. Another approach in \cite{cell4} suggests DL-based data augmentation techniques.

In \cite{new8}, a novel Long-Term Evolution (LTE) communication infrastructure-based environment sensing (LTE-CommSense) system is introduced, offering a non-intrusive, low-cost indoor positioning solution. Utilizing USRP B210, LTE data is gathered and analyzed using SVM and NN for detection of static and mobile occupants, achieving high classification accuracies. \cite{5gnew1} proposes a 5G-based navigation method employing ML-based software-defined receiver (SDR) for ToA estimation. Another 5G-based indoor positioning approach is suggested in \cite{5gnew2}, employing RSSI and DNN for localization. The comparison of these studies is listed in Table. \ref{wlancell}.
However, cellular signals being optimized for outdoor use, the signal undergoes degradation in indoors, hindering accuracy. Additionally, signal fluctuations, interference, and infrastructure reliance pose challenges for cellular-based IPSs, impacting their effectiveness and precision.

 \subsection{Ultra-wideband  (UWB)}

            UWB, unlike typical RFID systems, utilizes multiple frequency bands ranging from 3.1 GHz to 10.6 GHz \cite{uwb1}. It relies on ultra-short pulses narrower than 1 nm and a low duty cycle (1:1000), facilitating easy separation of original signals from multipath ones. UWB signals transmit for shorter durations than RFID signals and operate across a broad frequency range. Additionally, UWB tags require less power compared to conventional RF tags, and their interference with other RF signals is minimal due to signal type and spectrum differences.

            Various indoor localization techniques employing UWB data and neural networks are investigated in \cite{uwb2, uwb3,uwb4, uwbnew1}. These methods utilize channel impulse responses (CIR) from UWB indoor propagation measurements to construct fingerprint databases and estimate locations with positioning errors ranging from 0.081m to 2m. Some address NLoS challenges, exhibiting enhanced stability with updated databases. Others compare neural network models like MLP and RBF, exploiting UWB's high-time resolution for accurate positioning. Additionally, a CNN-based system in \cite{uwb8} accurately estimates distances, showing promise for indoor localization in unfamiliar environments with reduced complexity and low latency.
In \cite{uwb5}, a subterranean UWB-based fingerprinting radio localization system assisted by neural networks achieves improved stability and performance. Comparing two neural network models, MLP and RBF, yields positioning errors of 0.5m and 0.1524m, respectively. Another UWB-based strategy in \cite{uwb6} employs a back-propagating neural network to mitigate NLoS effects, while \cite{uwb8} introduces a CNN-based method for precise distance estimation without SNR estimation, thereby reducing latency and computational complexity. UWB signals for indoor localization offer accurate results with minimized multipath effects. TOA-based distance estimation between ANs and UWB tags is influenced by TOA errors and LoS conditions. A deep learning approach employing LSTM networks achieves a mean localization error of 7 cm, surpassing conventional methods \cite{rtt2,rtt3}. Recent research explores UWB for indoor localization, proposing a weighted indoor positioning algorithm that combines LSTM and DNN to analyze five UWB signal features, leading to enhanced positioning accuracy under severe NLoS conditions.

In \cite{uwbnew1}, a novel radar system combines UWB with ML, specifically, Multi-Class Support Vector Machine (MC-SVM), not only to localize targets but also to understand their locations effectively. Learning from extensive UWB signal data enables an evolving scheme to provide meaningful data for end-user appreciation. Experimental results affirm the effectiveness of the proposed algorithm. Another approach in \cite{uwbnew3} concentrates on enhancing UWB indoor localization using a Naive Bayes (NB) ML algorithm. Evaluation through Receiving Operating Curves (ROC) demonstrates significant enhancements in UWB localization accuracy, particularly in LoS and NLoS conditions. Table. \ref{wlancell} compares all these studies regarding complexity, scalability, and cost.

UWB-based IPSs encounter challenges like multipath interference, NLoS accuracy decline, and vulnerability to environmental conditions. Addressing these issues for reliable indoor positioning in dynamic settings is intricate. Interference from other wireless systems affects precision, while scalability and cost-efficiency challenges arise from requiring multiple UWB anchors. Advanced algorithms and hardware optimizations are crucial for synchronization and power consumption management.   

\begin{table*}
\renewcommand{\arraystretch}{1}	
		\caption{INDOOR DATASETS.}
		\begin{tabular}{ | m{3cm} | m{3cm}| m{1cm} |  m{1cm} |m{2cm}|m{4cm}|m{1cm}| } 
			\rowcolor{lightgray}
			\hline
% \vspace{.2cm}
				 Name & Technology  & Published Year  & Dimension & No. of Points & Scale & Paper \\ 
			\hline
   
CRAWDADa King \textit{ et al.}&RSSI&2008&1D& 146080, 6600&221 $m^2$, 1 Floor & \cite{dataset19}\\ \hline

   KIOS dataset &Wi-Fi RSSI&2013&1D&2100&560 $m^2$, 1 Floor & \cite{dataset20}\\ \hline
   
  UJIIndoorLoc & Wi-Fi & 2014 & 3D &19,937, 1111 & 108703 $m^2$&\cite{dataset1}\\ \hline
 
UJIIndoorLoc-Mag  & Magnetic & 2015 & 1D & 270, 11 & 260 $m^2$, 1 Floor &\cite{dataset21} \\ \hline

 KTH/RSS DATASET &RSSI&2016&1D&1689&400 $m^2$, Multiple Rooms&\cite{dataset13} \\ \hline

 Z. Tóth $\&$ J. Tamás & WLAN,BLE &2016 & 3D &1571& 3 Floors & \cite{dataset15}\\ \hline

 Barsocchi\textit{ et al.}\cite{dataset23}& Wi-Fi, Geo-Magnetic & 2016&1D &680, 460& 185 $m^2$, 3 Rooms, 4 Hallways & \cite{dataset23}\\ \hline

 Matterpport3D   &RGB-D Image&2017&3D&  194400& 90 Building-scale scenes & \cite{dataset4}\\ \hline

 Tampere database &Wi-Fi&2017& 3D &687, 3951 &22570 $m^2$, 4 Floors&\cite{dataset11} \\ \hline

Minho database   & Wi-Fi-RSSI&2017&1D&4973, 810& 1000 $m^2$ & \cite{dataset14}  \\ \hline

IPIN2017 Tutorial& Wi-Fi-RSSI & 2017&1D & 927, 702&School  corridor & \cite{dataset24}\\ \hline

InLoc  &RGB-D Image&2018&3D &277&Five Floors, 2 Buildings & \cite{dataset3} \\ \hline
    
Baronti\textit{ et al.} &BLE&2018&1D& 2598 &185 $m^2$, 7 Rooms &\cite{dataset7}\\ \hline

Bristol database &BLE - RSSI&2018&1D& 1571 & Multiple Houses &\cite{dataset8} \\ \hline

D. Sikeridis\textit{ et al.} & BLE & 2018&3D &-&  3 Floors &\cite{dataset22}\\ \hline

Mendoza-Silva\textit{ et al.} &Wi-Fi&2018&3D&576, 3120 & 308.4 $m^2$, 2 Floors &\cite{dataset9}\\ \hline

JUIndoorLoc &RSSI&2019&3D&23,904, 1460&882 $m^2$, 5 Floors& \cite{dataset12}\\ \hline

S. Sadowski \textit{ et al.}  & Zigbee, Bluetooth, Wi-Fi & 2020 & 1D &73, 32& 33 $m^2$-79 $m^2$, 3 Rooms  &\cite{dataset18} \\ \hline

F. Potortì \textit{ et al.}&Wi-Fi,Magnetic,Camera&2020& 3D&-& 6000 $m^2$, 3 Floors &\cite{dataset10}\\ \hline

SODIndoorLoc  database&Wi-Fi&2022&3D&  21,205,  2720  & 8000 $m^2$, 3 Builings,3 Floors & \cite{dataset6}\\ \hline

P. Pascacio\textit{ et al.}&BLE, RSSI&2022&1D& 178487 $\pm$ 85824 & 180 $m^2$  & \cite{dataset16}\\ \hline

TUJI1 Dataset &Wi-Fi&2023&1D&6752, 2147& Multiple Offices &\cite{dataset17}\\ \hline

		\end{tabular}
  \label{datasets}
	\end{table*}		

   \section{IPS Datasets }

   Indoor Positioning System (IPS) datasets are collections of data that have been gathered in indoor environments with the goal of developing and evaluating indoor localization and tracking algorithms. These datasets are invaluable for researchers, engineers, and developers working on improving indoor positioning technologies. Table \ref{datasets} contains details of various datasets collected using different technologies for developing IPSs.

The UJIIndoorLoc dataset \cite{dataset1} spans a university campus of approximately 110,000 $m^2$, created with 20 individuals and 25 Android devices, containing 19,937 training and 1,111 validation/test data points. UJIIndoorLoc-Mag \cite{dataset21} explores magnetic field variations with inertial sensor data. The JUIndoorLoc database \cite{dataset12} encompasses a five-floor building at Jadavpur University, with 25,364 samples collected from four Android devices, including 23,904 training and 1,460 testing samples, and 172 APs. SODIndoorLoc database \cite{dataset6} is a Wi-Fi-based dataset covering 8000 $m^2$ with three buildings (1-3 floors), including 105 pre-installed APs (56 single-band, 49 dual-band), 1802 points, 1630 RPs, 272 Testing Points (TPs), 23,925 samples, and three scenes: office, meeting, and seamless hall corridor. Other datasets include Baronti et al.'s indoor BLE dataset \cite{dataset7} for localization and tracking, the Bristol database \cite{dataset8}, and \cite{dataset22}, both BLE-based indoor positioning datasets supporting location-aware systems, crowd apps, and building management. The first one  covers multiple rooms in different houses and is collected using a wearable component package based on Texas Instruments CC2650 system-on-chip (SoC).
The latter collected from beacons carried by 46 individuals moving on three different floors of a university building.

The Library dataset \cite{dataset9} is comprised of Wi-Fi RSS data spanning 25 months from a university library measuring 308 $m^2$, featuring 576 training and 3,120 test samples collected via 620 APs. The IPIN 2019 Competition \cite{dataset10} assessed personal positioning systems across buildings, floors, and diverse scenarios, highlighting accurate positioning methods for real-world comparisons. The competition encompassed two adjacent buildings with outdoor and indoor areas totaling 1000 $m^2$ and 6000 $m^2$ respectively, spread over three floors with a path length exceeding 500 m. The Tampere database \cite{dataset11} gathered 4,648 fingerprints from 21 devices in a 22,570 $m^2$ university building, offering 687 training and 3,951 testing fingerprints. Additionally, the dataset \cite{dataset13} provides RSS data from indoor and outdoor environments covering approximately 400 $m^2$, including hallways, rooms, and an abandoned steel factory in Dortmund, Germany, recorded via an odometer. The Minho database \cite{dataset14}, collected at the University of Minho, Portugal, comprises 5,783 fingerprints, with 4,973 training samples, and covers around 1000 $m^2$ using 11 APs. Dataset \cite{dataset15} introduces a hybrid indoor positioning dataset  with WLAN, Bluetooth, and Magnetometer data. Unlike previous single-technology datasets, it covers 50$\%$ of a three-story building in which each floor covers an area of  1425 $m^2$ and the corridors and the hall cover approximately 465.75  $m^2$. Furthermore, BLE dataset \cite{dataset16} facilitates mobile-based indoor positioning investigations, encompassing calibration, NLoS ranging, and real office scenarios within a 10.76 m x 16.71 m area. It comprises 14 bookcases, 7 concrete columns, and 3 work sections with desks, chairs, and computers. Lastly, dataset \cite{dataset17} from UJI, Spain, presents various office and corridor settings with diverse wireless coverage, including Wi-Fi RSS, BLE RSS, accelerometer, and GNSS data, gathering approximately 9,000 annotated samples using GetSensorData 2.1, annotated with coordinates.

Raspberry Pi 3 Model Bs were utilized to establish a WiFi-based WLAN, utilizing onboard antennas in \cite{dataset18}. The CRAWDAD \cite{dataset19}, KIOS \cite{dataset20}, and IPIN 2017 \cite{dataset24} Tutorial datasets consist of RSS fingerprints of small areas with varying distances between location points. Dataset \cite{dataset3} is an image-based collection comprising 277 RGBD panoramic images from scanning two Washington University buildings with a Faro 3D scanner, each housing around 40M colored 3D points. Split into five scenes (DUC1, DUC2, CSE3, CSE4, CSE5) representing different floors, they are aligned to floor plan areas ranging from 23.5 to 185.8 $m^2$. Matterport3D \cite{dataset4} is another image-based dataset featuring 194,400 RGB-D images, obtained through 10,800 panoramas using a Matterport camera. It offers unique features like human-height viewpoints, global camera pose consistency, instance-level semantic segmentations, and data collected from private home living spaces, distinguishing it from previous datasets.

	\section{Future Directions}

 \begin{figure*}[t]
 \centering
     \includegraphics[width=0.6\textwidth]{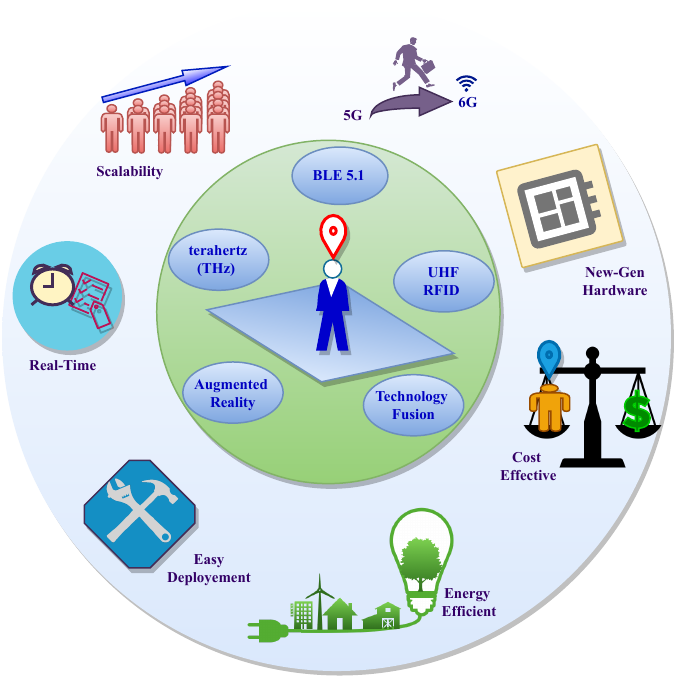}
     \caption{Future trends in IPSs.}
     \label{futuredia}
 \end{figure*}
IPSs are revolutionizing the way we navigate and interact with indoor spaces. As we move further into the future, the role and capabilities of indoor positioning systems are poised to become increasingly prominent and transformative. These systems offer the potential to enhance various aspects of our lives, from improving efficiency in industries to enhancing user experiences in retail, healthcare, and beyond. A visual depiction illustrating the forthcoming trends in IPSs through the utilization of diverse technologies is presented in Fig. \ref{futuredia}.

Future research into Wi-Fi fingerprint-based localization should explore the viability of CSI as a reliable method, leveraging its fine-grained characteristics for precise positioning. Incorporating vision into Wi-Fi localization offers expanded possibilities for richer location data. Ensuring accessibility for physically impaired individuals is crucial for inclusive indoor positioning services. Developing floor recognition for multi-level indoor localization enables seamless tracking across various floors. Optimizing deployment strategies is vital for creating practical Wi-Fi-based indoor positioning systems that fulfill diverse application needs. Large-scale location-based data mining holds promise for revealing valuable applications and insights. Integrating localization systems with 5G infrastructure must address challenges concerning power consumption and cost. Leveraging cooperative positioning techniques and data fusion from multiple 5G devices can enhance real-time localization accuracy. Key considerations for 5G-oriented systems include addressing NLoS scenarios and implementing robust methodologies for multi-path conditions and floor-level identifications. Integration of dead-reckoning techniques with visual place recognition can enhance pedestrian localization. While ensemble models and integrating NN-based methods with KF-based methods improve accuracy and reliability, careful deliberation is necessary due to complexity and latency concerns.

Challenges in WSN-based IPSs vary with applications, network scales, and environments, necessitating cost-effective solutions, diverse sensor integration, scalability, and computational complexity management. Balancing accuracy and cost-effectiveness, particularly in range-free techniques, is crucial. The future of UWB positioning entails hardware advancements and innovative algorithms for applications like multi-robot positioning and smart logistics. Integrating low-earth orbit satellites with UWB, GNSS, and inertial sensors expands indoor and outdoor navigation possibilities.

Ultrasonic-based IPSs need to tackle environmental challenges like noise and multipath propagation to boost location accuracy. Acoustic IPS research needs to prioritize real-time tracking, energy efficiency, multi-user localization, and robustness. ML algorithms and sensor fusion can further enhance performance. Optical IPSs should focus on accuracy, scalability, and high update rates. Integrating INS, GNSS, and magnetic sensors can improve system performance. Cost-effective solutions are essential for widespread adoption. In mmWave IPSs, advanced hardware and scalable algorithms are crucial for faster localization and multi-AP management. Integrating ML presents challenges that require reduced reliance on extensive training datasets. Combining mmWave radar with camera-based systems and adapting IEEE 802.11ay for passive sensing hold potential for advanced applications.

The future of indoor and outdoor localization hinges on leveraging ML tools to broaden their applicability. ML effectively tackles noisy environments and NLoS conditions, improving data analysis. Key research focuses on floor detection in multi-storey buildings, demanding 3D localization techniques empowered by ML-based 3D location estimation. Neural networks aid in feature extraction, ensuring robust results for tasks like human detection. ML simplifies localization, accommodating diverse scenarios with suitable hardware. Cost-effective deployment enables large-scale data mining, enabling applications like indoor landmark discovery. CSI's robustness in fingerprinting may supplant RSSI, while integrating vision, sound detection, and motion sensors enhances localization, especially for the physically challenged. Despite challenges, ML-based indoor positioning gains traction, meeting global demands. The evolving landscape demands innovative indoor positioning systems, fostering competition and advancement in the field.

   \section{Conclusion  }

Human localization plays a pivotal role in various applications, from monitoring elderly health to safety operations. Consequently, extensive research has been dedicated to developing efficient indoor positioning systems. Hence, this paper  explores  different machine learning-based indoor localization approaches discussed in the literature along with their performance comparisons. Through the assessment and comparison of diverse machine learning methods, this study aims to contribute to the advancement of indoor positioning systems and their application in diverse domains. The survey also encompasses publicly available datasets pertinent to indoor localization and concludes by outlining future trends associated with each sensing technology.

We foresee numerous research directions leveraging ML tools in both indoor and outdoor localization techniques to enhance their widespread deployability. ML tools help to address the noisy environment in NLoS conditions and enable more efficient data analysis. Floor detection for multi-storey buildings is a crucial research direction in  indoor localization, necessitating a transition to 3D localization techniques. ML-based 3D location estimation  employs neural networks to extract relevant features and predict environmental data efficiently and more accurately. Consistency and reliability are essential for indoor localization applications, where ML methods notably impact human detection, intruder identification, and other tasks. By utilizing ML tools, data analysis becomes more reliable, even in complex environments. Traditional approaches perform differently indoors and outdoors due to environmental variations, necessitating adaptable solutions. Addressing cost concerns, recent advancements enable high localization performance at reduced deployment costs. The widespread deployment of these systems facilitates large-scale location-based data mining, supporting applications like indoor landmark discovery, queue detection, and geofencing.
		
			\par Considering different data types, CSI is likely to replace RSSI due to its higher robustness and performance in fingerprinting.  However, smartphone limitations hinder CSI data collection, necessitating specialized Wi-Fi-enabled infrastructure. Additionally, Channel impulse response (CIR) and Signal Eigenvector are also drawing attention for better indoor localization. Integrating vision with fingerprinting offers promising avenues, while combining various technologies like sound detection, inertial sensors, and motion sensors can improve localization, particularly for assisting physically disabled individuals \cite{intro12}.

Establishing a dependable indoor positioning system remains challenging due to environmental factors. Integrating ML into the system enhances localization reliability. Despite cost and complexity, indoor positioning systems are widely adopted worldwide. Diverse applications necessitate innovative systems, fostering competitiveness. As a comprehensive survey covering all the subdomains is currently absent, this paper serves as a valuable resource for referencing state-of-the-art approaches within this emerging realm. The principal motivation driving this endeavor is to offer insights into this intricate domain and to contribute to research challenges associated with it. Its aim is to provide essential references for readers to gain knowledge and contribute further in this field.

			\ifCLASSOPTIONcaptionsoff
			\newpage
			\fi

	\begin{IEEEbiography}[{\includegraphics[width=1in,height=1.25in,clip,keepaspectratio]{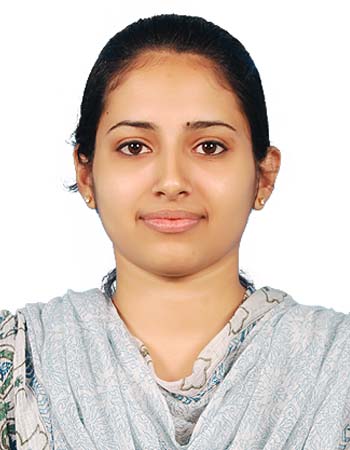}}]{Amala Sonny}  received the B.Tech. degree
in electronics and communication engineering
from the TKM college of engineering, Kollam, India, in 2014, and the M.Tech. degree from the National Institute of Technology Calicut, Calicut, India, in 2017.
She is pursuing  the PhD degree with the Department of EE,  Indian Institute
of Technology Hyderabad, Hyderabad, India.  She was  a visiting researcher  at the Department of ICT, University of Agder, Kristiansand, Norway. Her research interests include  Indoor and outdoor localization, machine learning, and wireless communication. 
\end{IEEEbiography}

% \begin{IEEEbiography}[{\includegraphics[width=1in,height=1.25in,clip,keepaspectratio]{amuru2.png}}]
%    { SaiDhiraj Amuru (Senior   Member, IEEE)} (S’12–M’15) received the
% B.Tech. degree in electrical engineering from
% IIT Madras, Chennai, India, in 2009, and the Ph.D.
% degree in electrical and computer engineering from
% Virginia Tech in 2015. His Ph.D. advisor at Virginia
% Tech was Dr. R. M. Buehrer. From 2009 to 2011,
% he was with Qualcomm, India, as a Modem Engineer. He visited the Networks, Economics, Communication Systems, Informatics, and Multimedia
% Research Laboratory, University of California at
% Los Angeles, Los Angeles, CA, USA, in 2014. He is
% currently a Chief Engineer with the Samsung Research and Development
% Institute, Bangalore. His research interests include cognitive radio, statistical
% signal processing, and online learning.
% \end{IEEEbiography}

\begin{IEEEbiography}[{\includegraphics[width=1in,height=1.25in,clip,keepaspectratio]{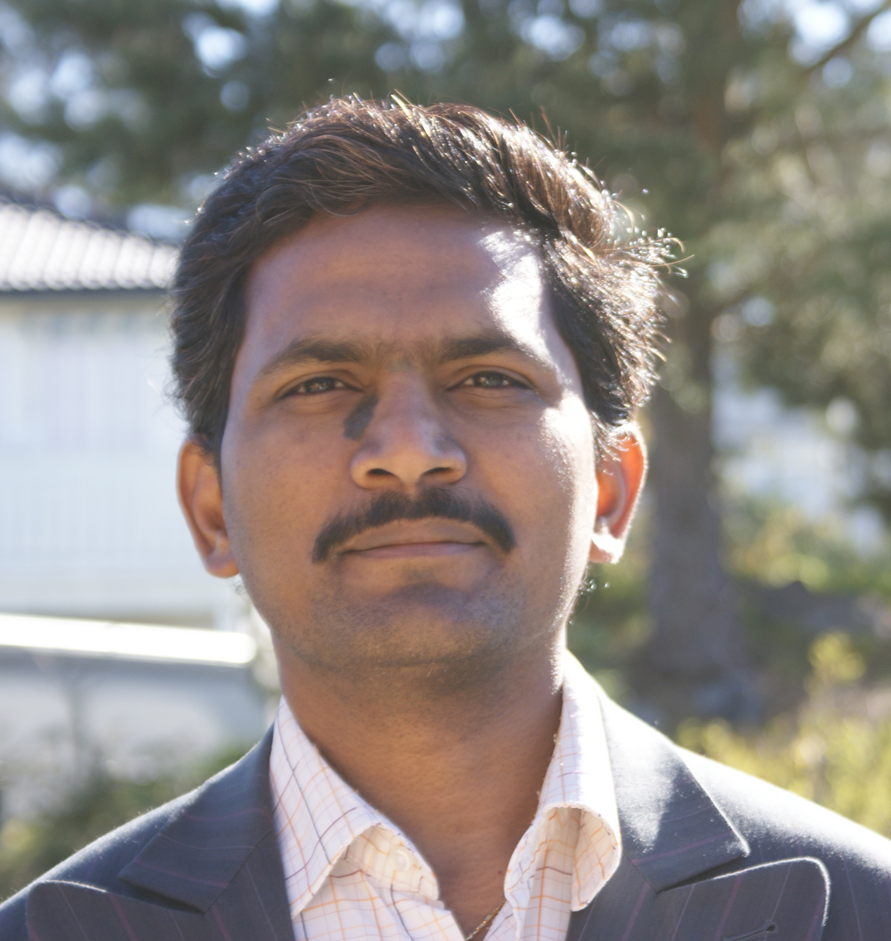}}]{Linga Reddy Cenkeramaddi (Senior   Member, IEEE)} received the master’s degree in electrical engineering from the Indian Institute of Technology Delhi (IIT Delhi), New Delhi, India, in 2004, and the Ph.D. degree in electrical engineering from the Norwegian University of Science and Technology (NTNU), Trondheim, Norway, in 2011.

He worked at Texas Instruments on mixed-signal circuit design before joining the Ph.D. Program at NTNU. After finishing his Ph.D. degree, he worked on radiation imaging for an atmosphere–space interaction monitor (ASIM mission to the International Space Station) at the University of Bergen, Bergen, Norway, from 2010 to 2012. He is currently the Leader of the Autonomous and Cyber-Physical Systems (ACPS) Research Group and a Professor at the University of Agder, Grimstad,  Norway. He has co-authored over 150 research publications that have been published in prestigious international journals and standard conferences in the research areas Internet of Things (IoT), Cyber-Physical systems, Autonomous systems, Robotics and Automation involving advanced sensor systems, Computer vision, Thermal imaging, LiDAR imaging, Radar imaging, wireless sensor networks, smart electronic systems, advanced machine learning techniques, connected autonomous systems including drones/unmanned aerial vehicles (UAVs), unmanned ground vehicles (UGVs), unmanned underwater systems (UUSs), 5G- (and beyond) enabled autonomous vehicles, and Socio-Technical Systems like urban transportation systems, Smart Agriculture and Smart Cities. He is also quite active in medical imaging. Several of his master's students won the best master thesis award in information and communication technology (ICT). % Dr. Cenkeramaddi is also a senior member of IEEE, a member of ACM, and a member of the editorial boards of various international journals and the technical program committees of several IEEE conferences. He is the Principal Investigator and a Co-Principal Investigator of many research grants from the Norwegian Research Council.
\end{IEEEbiography}
\begin{IEEEbiography}[{\includegraphics[width=1in,height=1.25in,clip,keepaspectratio]{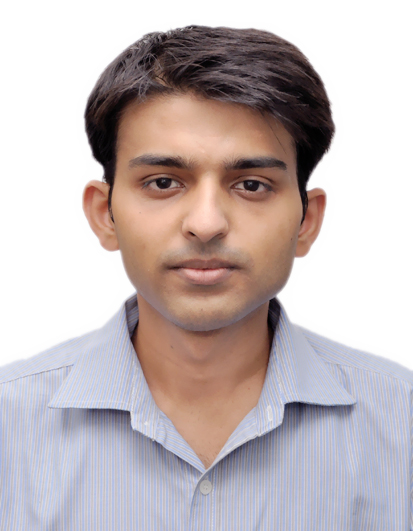}}]{Abhinav Kumar (Senior   Member, IEEE) } received the B.Tech. degree in electrical engineering, and the M.Tech. degree in information and communication technology, and the Ph.D. degree in electrical engineering from the Indian Institute of Technology (IIT) Delhi, in 2009, 2009, and 2013, respectively. He was a Research Associate with IIT Delhi in 2013. From 2013 to 2014, he was a Postdoctoral Fellow at the University of Waterloo, Canada. Since 2014, he has been with IIT Hyderabad, India, where he is currently an Associate Professor. His research interests are in different aspects of wireless communications and networking.
\end{IEEEbiography}

  \end{document}